\documentclass[fleqn,10pt,dvipsnames]{wlscirep}
\usepackage[utf8]{inputenc}
\usepackage[T1]{fontenc}

\usepackage{rotating}
\usepackage{threeparttable}
\usepackage{float}

\title{PhagoStat a scalable and interpretable end to end framework for efficient quantification of cell phagocytosis in neurodegenerative disease studies}

\author[1]{Mehdi Ounissi}
\author[2,3]{Morwena Latouche}
\author[1,*]{Daniel Racoceanu}

\affil[1]{Sorbonne University, CNRS, Inserm, AP-HP, Inria, Paris Brain Institute - ICM, Paris, 75013, France}
\affil[2]{Sorbonne Université, Inserm, CNRS, AP-HP, Institut du Cerveau, ICM, F-75013 Paris, France}
\affil[3]{PSL Research university, EPHE, Paris, France}

\affil[*]{Corresponding author: daniel.racoceanu@sorbonne-universite.fr}


\begin{abstract}

Quantifying the phagocytosis of dynamic, unstained cells is essential for evaluating neurodegenerative diseases. However, measuring rapid cell interactions and distinguishing cells from background make this task very challenging when processing time-lapse phase-contrast video microscopy. In this study, we introduce an end-to-end, scalable, and versatile real-time framework for quantifying and analyzing phagocytic activity. Our proposed pipeline is able to process large data-sets and includes a data quality verification module to counteract potential perturbations such as microscope movements and frame blurring. We also propose an explainable cell segmentation module to improve the interpretability of deep learning methods compared to black-box algorithms. This includes two interpretable deep learning capabilities: visual explanation and model simplification. We demonstrate that interpretability in deep learning is not the opposite of high performance, by additionally providing essential deep learning algorithm optimization insights and solutions. Besides, incorporating interpretable modules results in an efficient architecture design and optimized execution time. We apply this pipeline to quantify and analyze microglial cell phagocytosis in frontotemporal dementia (FTD) and obtain statistically reliable results showing that FTD mutant cells are larger and more aggressive than control cells. The method has been tested and validated on several public benchmarks by generating state-of-the art performances. To stimulate translational approaches and future studies, we release an open-source end-to-end pipeline and a unique microglial cells phagocytosis dataset for immune system characterization in neurodegenerative diseases research. This pipeline and the associated dataset will consistently crystallize future advances in this field, promoting the development of efficient and effective interpretable algorithms dedicated to the critical domain of neurodegenerative diseases' characterization. \href{https://github.com/ounissimehdi/PhagoStat}{GitHub}, \href{https://zenodo.org/records/10803492}{Dataset}.

\end{abstract}

\begin{document}

\flushbottom
\maketitle

\thispagestyle{empty}

\noindent %
\section*{Introduction} \label{sec:introduction}

The latest advances in physics, especially in high-throughput microscopy, along with computer-aided analysis, are opening up a new frontier in fundamental cellular understanding in biology. The progress made in this field is due to the automation of formerly labor-intensive tasks, such as identifying cells, counting them, tracking their movements, and profiling their characteristics~\cite{analysis_strategies,methods_tracking,computers_phenotypes}.

These developments have important implications for numerous biological experiments that explore the dynamics and the behavior of immune cells. In particular, the process of phagocytosis, in which microglial cells engulf and eliminate protein deposits or aggregates, has garnered attention in the field of neurodegenerative diseases~\cite{A,B,C,D,E, NeuRegenerate}. A deeper understanding of this phenomenon is essential for elucidating the complex mechanisms underlying these disorders and their progression. Consequently, there is a growing demand for more precise and accurate quantitative methods to advance the field, as they can offer invaluable insights into the interplay between microglial cells and protein aggregates, ultimately contributing to the development of novel therapeutic strategies and interventions for neurodegenerative diseases.

While traditional image processing methods have been used in microscopy to detect cells~\cite{buggenthin2013automatic}, they often struggle to accurately detect unstained cells, measure rapid cell interactions, and distinguish cells from their background in a complex setting. To address these challenges, innovative approaches that utilize the latest advances in computer vision (CV) and deep learning (DL) are required~\cite{survey_deep, DeepAnalysisSurvey}.

Fortunately, DL has brought substantial improvements to cell segmentation, offering several advanced models including U-Net, Mask R-CNN, DeepLabv3+, Stardist, and Cellpose, which have been widely adopted for various segmentation tasks~\cite{unet,mask_rcnn, deeplabv3plus2018, starconex, cellpose, Arbelle}.

However, the opaqueness of black-box algorithms, presents a major obstacle to their adoption in clinical settings. Therefore, it is important to reveal the inner workings of DL models and promote transparency~\cite{VANDERVELDEN2022102470, xai_arreta} to build trust in using advanced technology. Consequently, the development of interpretable DL-based tools can encourage their adoption, potentially leading to further discoveries in these fields.

Despite the potential of DL-based approaches for quantifying cell phagocytosis, to our knowledge, there is currently no existing ready-made open-source solution capable of effectively handling this task under real-world conditions. To address this, we introduce PhagoStat, a scalable pipeline utilizing interpretable DL and open-source packages to enhance our understanding of cell phagocytosis process. This pipeline integrates DL precision with explainable artificial intelligence (XAI) transparency, emphasizing interpretability in cell biology research. PhagoStat streamlines essential cellular feature extraction, providing a comprehensive, accessible solution for researchers and advancing our knowledge of dynamic cellular processes, particularly phagocytosis.

\begin{figure}[ht]
    \centering
    \includegraphics[width=\textwidth]{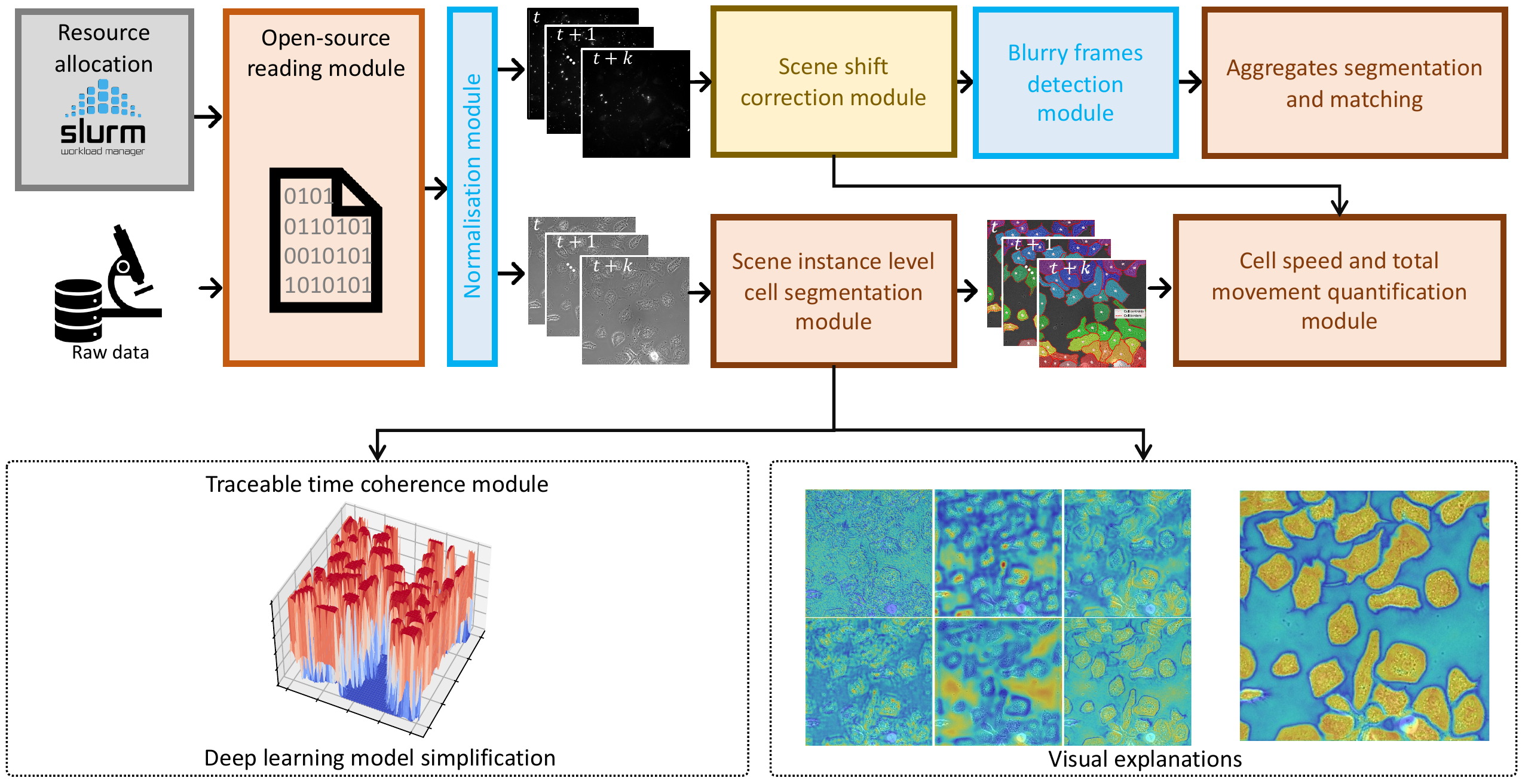}
    \caption{\textbf{PhagoStat: A comprehensive end-to-end pipeline for quantifying microglial cell phagocytosis in the context of frontotemporal dementia (FTD).} The PhagoStat pipeline is a fully operational system comprised of the following stages: (i) efficient loading of raw data (Fig.\ref{fig2}.b), (ii) applying data quality checks and quantifying aggregates over time (Fig \ref{fig3}.c), and (iii) performing cell instance segmentation using an interpretable deep learning (IDL) approach (Fig.\ref{fig5}, which incorporates Fig.\ref{fig4}.c). This comprehensive pipeline streamlines the analysis process and facilitates accurate and reliable results for researchers working with microglial cell phagocytosis data.}
    \label{fig1}
\end{figure}

The principal contributions of our work can be outlined as follows:

\begin{enumerate}

    \item The PhagoStat pipeline, illustrated in Fig.~\ref{fig1}, is a flexible framework comprising of several interconnected and interchangeable modules designed to analyze unstained cells overlaid on fluorescent aggregates. It handles data-sequence loading and normalization, video-based registration and frame alignment, noise detection and removal, aggregate and cellular quantification, and statistical reporting. By streamlining the analysis process, these modules ensure precise results, preserve data integrity, leverage interpretable DL for cellular segmentation, and provide clear statistical insights into cell activity under different experimental conditions.

    \item We showcase a comprehensive use case of the PhagoStat pipeline applied to microglial cells. Our analysis yields statistically significant results, uncovering a crucial finding: Frontotemporal Dementia (FTD) mutant cells display both larger size and increased activity in comparison to their control counterparts (wild-type: WT), refer to section~\ref{microglial-use-case}. This important discovery demonstrates the potential of the PhagoStat pipeline in advancing our understanding of neurodegenerative diseases and promoting further research in this critical field.
    
    \item We are releasing a large dataset that focuses on the phagocytosis of microglial cells. The dataset includes ten videos captured over a 7-hour period. Each video containing 20 simultaneous scenes, with both WT and FTD mutants cells in the context of FTD. The aggregates are fluorescent, while the cells are unstained. The purpose of releasing this dataset is to facilitate further understanding and application of our pipeline. The dataset comprises 36,496 cell images and 1,306,131 cell masks, along with 36,496 aggregate images and 1,553,036 aggregate masks, promoting further research into microglial cell phagocytosis.
    
\end{enumerate}

\section{Results} \label{sec:results}

\subsection{Phagocytosis quantification pipeline}

The pipeline consists of several interconnected modules, each tailored to perform specific tasks effectively.

Firstly, we introduce the data-efficient loading and normalization module (refer to~\ref{data_module}), which ensures streamlined data handling and preprocessing. This module is crucial for reducing the computational burden and facilitating subsequent analysis.

Secondly, we elaborate on the spatiotemporal frame registration process, which is accompanied by a data quality check (refer to~\ref{frame_registration}). This step incorporates scene shift correction and blurry frame detection, ensuring the accuracy and reliability of the acquired data while maintaining spatial and temporal alignment across frames.

Thirdly, We introduce the cellular and aggregate quantification modules (refer to~\ref{cell_seg}) designed to offer a comprehensive analysis of cellular properties and interactions. The first module employs instance-level interpretable segmentation techniques, effectively extracting and analyzing cellular properties from unstained cell images. This approach allows for an in-depth examination of individual cell behavior and morphology. Concurrently, a second module utilizes advanced image processing techniques to facilitate accurate aggregate segmentation and matching from fluorescent aggregate images. These techniques provide insights into the complex interactions between cells and aggregates, revealing crucial biological and morphological features.

Finally, the pipeline integrates the information obtained from both aggregates and cells. By quantifying phagocytosis, this module provides a comprehensive projection of the entire framework, showcasing the applicability of PhagoStat pipeline in the phagocytosis research contexts.

\subsubsection{Data efficient loading and normalization}\label{data_module}
We would like to emphasize that this module is not designed exclusively for a specific microscope or proprietary software. Instead, it presents a versatile and general approach that can be readily applied to any microscope system. This use case serves to demonstrate the effectiveness of the module, which was implemented using open-source packages, with further details provided in the Methods section. Our findings illustrate that our approach is at least two times faster than commonly used proprietary software while requiring only 1/8 of the hardware resources, as shown in Fig.~\ref{fig2}.c.

Following a thorough evaluation, we observed that the combination of percentile normalization (tailored to each sequence, referred to as "global normalization") and cumulative histogram distribution matching (tailored to each image, referred to as "local normalization") is highly effective in reducing data variability. This dual normalization strategy enhances the performance of DL segmentation model, resulting in up to a 10\% improvement in the dice coefficient. Moreover, we integrated this normalization approach with the raw data readout module, capitalizing on its parallelism scheme. This integration substantially reduces the time cost associated with various processes, such as reading raw data, normalizing data, and saving normalized data, as opposed to adding a separate normalization module after data readout (e.g., reading raw data, saving it in an open format, loading saved data, normalizing data, and saving normalized data).

\subsubsection{Data quality check for scene shift correction and blurry
frames rejection}\label{frame_registration}

During our analysis of high-content video-microscopy recordings captured over extended periods, we observed the presence of several unavoidable hardware-related acquisition faults or artifacts. One such artifact was the unintended shaking of the microscope along the x and y axes. These imperfections, although localized to specific frames, had a detrimental impact on the overall sequence quality. Consequently, these artifacts can potentially compromise the accuracy and reliability of subsequent data analysis and interpretation, underscoring the importance of addressing such issues in a systematic manner during the data processing stage.

In our study, we aimed to investigate how external disturbances affect the performance of the microscope camera sensor. To this end, we recorded 20 simultaneous scenes, each containing two channels (cells:non-fluorescent and aggregates:fluorescent), with a frequency of 1 frame every 2 minutes, over a period of 7 hours without any interruption. During the recording process, we estimated that the microscope camera sensor had only 6 seconds (2 min divided by 20 scenes) to cycle and stabilize along the \textit{x} and \textit{y} axes from frame \textit{n} to frame \textit{n+1} to capture pixel intensities and write them to a local disk.

However, external disturbances, such as mechanical vibrations, lens getting out of focus, can cause the microscope camera sensor to deviate from its normal performance, further reducing the time-response gap. Likely, such external disturbances occur at least once during the 7h non-stop sessions. When this happened, it affected the quality of 1 to 10 consecutive frames.

In order to counteract the potential influence of external disturbances on our analysis, we developed a registration-based module specifically designed to align and stabilize the frames. This module effectively mitigates any potential deviations caused by external factors, ensuring the accuracy and reliability of the subsequent data analysis.

It is important to note that in the aggregate channel, the majority of pixels belong to the background. This predominance of low-intensity pixels (i.e., the background) in comparison to the high-intensity pixels (i.e., the aggregate) presents a challenge when it comes to registration (offset correction).

Given the explainable nature of our pipeline, we opted against using black-box registration approaches based on DL~\cite{registration_survey}. Instead, we chose to employ the Scale-Invariant Feature Transform (SIFT) algorithm~\cite{SIFT} as our registration method. SIFT is not only fully explainable and mature, but it has also become publicly available since the expiration of its patent in March 2020. This combination of explainability and accessibility makes SIFT an ideal choice for our pipeline.

Although the application of SIFT proved sufficient in correcting the shift, as demonstrated in Fig.~\ref{fig6}.a and in extended data \ref{extended} Table.\ref{tab:reg}, we observed a statistically significant directional bias in the shift correction. As a result, we concluded that an unbiased approach was necessary to address the registration problem effectively. To this end, we proposed a generalized version of the Enhanced Correlation Coefficient maximization approach (ECC)\cite{correlation_reg}, which we have termed Cascade ECC (CECC), as illustrated in Fig.~\ref{fig3}.a.

CECC offers an unbiased solution, ensuring consistent registration performance irrespective of the shift direction. Our approach achieved sub-pixel precision for shift margins up to $\pm20\%$ of the image size (i.e., $\pm400px$ for $2048\times2048px$ frames). To provide context for these results, the largest unwanted shift observed in our study was approximately 5\% of the image size. This demonstrates the robustness of CECC, which offers a 15\% margin in the context of the worst-case scenario observed. It is worth noting that there is a significant difference in registration speed between SIFT, with an average of approximately 3 seconds per frame, and CECC (n=5), with an average of approximately 8.7 seconds per frame (refer to Fig.~\ref{fig6}.b). This discrepancy can be attributed to the maturity and optimization of SIFT, as compared to the proposed CECC. We anticipate that this gap will eventually be reduced through the contributions of the open-source community.

An other issue worth mentioning when we explored the data is that the microscope lens can get getting out of focus because of physical vibration. This introduce some blurry frames into the scene, which unnaturally amplify the size of aggregates, thus, biasing the phagocytic quantification.

We addressed this issue, by including a blur detection module, that uses image processing to detect 
the loss of details in images. Then discard them from the stack.

To streamline the data quality check process we combining the CECC-based scene shift correction and the blurry frames detection module (see Fig.~\ref{fig3}.b).

This gives the user full traceability over the data quality, making the definition of objective criteria possible. For example, the maximum tolerated shift along either axis x and y can not be more than 50 pixels at any frame, and the maximum tolerated blurry frames in a given scene can not be more than 5\%. The data quality check module plays a crucial role in detecting (i.e., blurry frames), correcting (i.e., scene shift) and objectively quantifying the severity of hardware/human errors in real-world conditions.

\subsubsection{Cellular and aggregates quantification}\label{cell_seg}

\textit{Cellular quantification:} The complex and challenging nature of cell instance segmentation is a important issue in the field of biological image analysis. This problem involves detecting cells with intricate shapes and various morphology, which has prompted the development of numerous DL-based approaches to address it~\cite{Turaga2010, unet, deepwatershed, lstm_unet, Moen2019, starconex,cellpose}. While these techniques have proven to be effective, they fail to exploit the temporal information that may be critical for accurately segmenting cells with irregular shapes, such as microglia, which are known for their hyperactivity.

Incorporating spatiotemporal approaches to cell instance segmentation has the potential to greatly improve the accuracy of detection by considering cell movement over multiple frames~\cite{Arbelle,Liang}. This approach can help correct the detected cell masks by introducing time coherency~\cite{Liang}, allowing for a more precise identification of individual cells and their boundaries.

Additionally, it should be noted that contemporary high-performing methods like TrackMate\cite{TrackMate}, designed as plugins for Fiji, are best suited for light usage and not ideally suited for the automated analysis of extensive datasets, requiring significant computational power.

To address these challenges, we propose a generic scheme for robust cell instance segmentation. Our method consists of three primary components: (i) a cell semantic segmentation module, which provides accurate semantic masks for distinguishing cells from the background; (ii) a time-series coherence module that leverages cell movement over time to enhance instance separation; and (iii) a post-processing and refinement step that fuses the instance and semantic information to generate a more accurate representation of cell boundaries.

To substantiate the added value of incorporating temporal information in cell segmentation, we constructed a DL only approach that leverages UNets~\cite{unet, attunet, bionet} (UNet, AttUNet and BioNet) as a semantic segmentation module, complemented by the integration of long short-term memory (LSTM)\cite{lstm_serv} as a time series coherence module (see Fig.~\ref{fig4}.a). Furthermore, we utilized post-processing techniques to fuse the previous signals\cite{watershed1, watershed3}, ensuring complete separation between cells (see Fig.~\ref{fig4}.b).

As evidenced by Fig.~\ref{fig6}.c, our approach, which combines DL and temporal information (AttUNet-LSTM, UNet-LSTM), demonstrates improved performance compared to state-of-the-art methods, such as Cellpose and Stardist. However, we observed performance variability when using the BioNet-LSTM (included as a fail case). Upon further investigation, we identified the cause to be rooted in the unique architecture of BioNet-LSTM, which employs an internal recursive mechanism that drastically over-fit the feature maps. This limitation consequently hampers its potential when connected to a time series module, thereby impacting its overall performance. Nonetheless, our findings highlight the importance of incorporating temporal information in cell segmentation approaches to achieve more accurate and reliable outcomes in the analysis of cellular dynamics.

However, the proposed UNets+LSTM approaches, being solely based on DL (DL-only), do not offer interpretability. Despite their superior performance compared to state-of-the-art methods, we consider this lack of interpretability insufficient. We believe that interpretability offers a substantial added value to these approaches. To demonstrate that interpretability does not adversely affect performance, we first initiated the process of "whitening the black box" (i.e., the DL-only approach) to establish that it is not only feasible but also quantitatively comparable to its counterparts (DL-only, Cellpose, and Stardist).

We explored explainable methods~\cite{xai_arreta, xai_visual, attunet} and determined that post-hoc feature visualization, represented in the form of heat maps with colors ranging from red for essential features to blue for less critical ones, would serve as a reliable means of gaining insights into the workings of the DL-only approach. This approach would allow for the potential simplification of the process, if deemed necessary, without compromising the effectiveness of the cell segmentation methods.

We generated heat maps for unseen images and attempted to interpret them for all components of the DL-only (UNets+LSTM) approach. Our analysis yielded the following results: for UNets, a substantial amount of unnecessary redundancy in feature maps was observed; LSTM features focused on the less mobile internal parts of cells while utilizing their movement (temporal information) to smooth out cell boundaries for improved separation. From these observations, we concluded that the default number of trainable parameters in UNets was unnecessarily high, and the LSTM effectively harnessed temporal information in a manner that enhanced the overall segmentation process.

After training on a supervised dataset, our approach optimizes the models quantitatively, by evaluating their efficiency in the feature space, without depending on annotated test sets like nnU-Net~\cite{nnUNET} and NAS~\cite{NAS}. This evaluation is crucial as nnU-Net and NAS, despite their strengths, rely on such sets, potentially leading to biases with regards to data annotation or selection. This data-independent method excels especially in situations where unbiased annotations are resource-intensive, enabling optimization on all existing, non-annotated data. By quantitatively monitoring performance metrics, we fine-tuned the trainable parameters of U-Nets, achieving a measurable 7-fold decrease in model size, which is documented in Fig.~\ref{fig6}.e. Also, we provided relevant details along with a use-case where we balance feature map quality using quantitative metrics and execution time in Unet models (refer to the methods section: 'Automated DL model optimization using feature maps' and to Fig.~\ref{sup_fig_features}). The effectiveness of our method in reducing over-optimization and improving generalization, provides a stark contrast to the limitations seen in nnU-Net and NAS.

Inspired by the LSTM strategy, we devised an image processing-based algorithm called the Traceable Time Coherence Module (TTCM). This algorithm takes into account a time window of cell mask predictions and assigns probabilities ranging from 1 for static cell parts to 0 for moving parts, in a manner that enables cell separation. By incorporating TTCM, we aimed to enhance the overall segmentation process by leveraging temporal information in a similar way to the LSTM approach (refer to Fig.~\ref{fig4}.b and Fig.~\ref{fig4}c).

The outcome of our efforts was a more compact UNets in comparison to the DL-only baseline. For simplicity, we will refer to these compact versions as UNets(XAI). The visual interpretation-guided optimization was complemented with a visual explanation module to enhance the trustworthiness of UNets. This was achieved by systematically generating heat maps to be presented to biologists and clinicians, illustrating the intermediate steps involved in obtaining the cell mask (refer to Fig.~\ref{fig4}.c).

Additionally, the TTCM exhibited unexpectedly advantageous properties compared to the LSTM approach. Notably, it eliminated the need for important hardware requirements during both training and inference phases, required no training, and provided a flexible time window parameter that can be adjusted as needed—a feature particularly beneficial in research environments. For example, larger time windows can be applied for slow-moving cells, while smaller windows may be more suitable for faster cells.

In terms of performance evaluation, we employed the mean intersection over union (mIoU) metric for each image. This involved comparing the sum of IoU scores of the predicted cell masks with the ground truth masks and subsequently dividing the result by the cell count. This process was carried out for each image in the test set (n=165) using a 5-fold testing approach. The mean and standard deviation were computed for state-of-the-art methods: Cellpose, Stardist, DL-only: Bionet-LSTM, AttUNet-LSTM, UNet-LSTM, and IDL: Bionet(XAI), AttUNet(XAI), UNet(XAI) approaches.

It is important to highlight that in Fig.~\ref{fig6}.d, we assessed the inference speed of all considered approaches. For example, Cellpose relies on intensive post-processing to transform vector gradient representation into labeled cell masks, which results in a slower performance compared to other methods.

Additionally, the mIoU metric is well-suited for the comprehensive evaluation of cell detection and segmentation quality without requiring a threshold. However, in the literature~\cite{starconex,cellpose}, average precision (AP) is defined as:\\ \mbox{$AP={(TP + TN)}*{(TP+TN+FP+FN)}^{-1}$}, with $TN=0$. We contend that this metric is, in fact, the accuracy metric, which differs from the standard AP metric employed in object detection problems, defined as the area under the precision/recall curve. Indeed, accuracy has been used in previous studies~\cite{starconex,cellpose} to evaluate detection quality.

To maintain continuity with the metrics previously used in Cellpose~\cite{cellpose} and Stardist~\cite{starconex}, we included Fig.\ref{fig6}.f, where we applied the same accuracy metric to compare our best-performing mIoU approach – 'AttUNet(XAI)' – with state-of-the-art methods. We utilized different IoU thresholds and computed the mean and standard deviation values over 5-fold cross-validation and testing. Our method outperforms both Stardist and Cellpose when $IoU>0.8$. This result demonstrates that our approach is well-suited for fast, efficient, and high-quality instance-level cell segmentation. For a quantitative/qualitative comparison, please refer to the extended data\ref{extended}: Tab.\ref{tab:seg} and Fig.\ref{sup_fig_3}.

It is worth mentioning that we employed the default configurations of state-of-the-art methods (Cellpose and Stardist)~\cite{starconex,cellpose} as described in their respective publications, utilizing their provided source code. We observed a longer training time for these methods compared to our DL-only and IDL approaches. Specifically, Cellpose requires 500 epochs, and Stardist necessitates 400 epochs, while our IDL approach takes only 20 epochs, and the DL-only method takes 40 epochs. This observation highlights the efficiency of our proposed methods in terms of training time.

The key takeaways from our results are as follows:
\begin{itemize}
    \item Fig.~\ref{fig6}.c demonstrates that incorporating temporal information enhances the performance of segmentation methods, and interpretability can be achieved without sacrificing performance.
    \item Fig.\ref{fig6}.d and Fig.\ref{fig6}.e confirm that interpretability offers valuable optimization properties, such as improved processing speed and reduced hardware requirements.
\end{itemize}

Some unexpected results emerged when we applied feature visualization to UNet and AttUNet in the context of cell segmentation. The UNet model employs an encoder to extract background features, focusing on the mid-section 'bottleneck' of the model. Subsequently, the decoder disregards the background and utilizes skip connections to generate segmentation masks. This observation can be attributed to the relative complexity of modeling foreground texture, which exhibits a high degree of variation, as opposed to the uniform background. By interpreting the heat maps, we confirmed that UNet adopts a less demanding approach, concentrating on background feature extraction and then estimating the complementary mask to produce cell masks. In contrast, the AttUNet model incorporates an attention mechanism~\cite{attention, attunet} in addition to the UNet structure. Feature visualizations revealed that the attention mechanism actively alters the learning behavior of the model  (compared to UNet), directing its focus more towards cell features, unlike the UNet model. Interestingly, even when trained on the same dataset, UNet and AttUNet models can tackle the same task and achieve comparable segmentation performance while learning distinct sets of features (see extended data~\ref{extended}: Fig.~\ref{sup_fig_1}).\newline

\textit{Assessing Phagostat's segmentation generalization capability:} 
In order to benchmark our approach against current and future approaches, we trained and subsequently submitted the AttUnet(XAI) model for evaluation on the CTC datasets, which include Fluo-N2DL-HeLa, Fluo-N2DH-GOWT1, DIC-C2DH-HeLa, Fluo-N2DH-SIM+, and PhC-C2DH-U373. This assessment focused on segmentation tasks, as detailed in Tab.\ref{tab:TESTCTCseg}. For this process, both available training sequences along with their respective ground truth masks were utilized. The training was conducted from scratch, following an 80-20 split for training and validation, over a span of up to 200 epochs. Notably, the model's effectiveness was measured by the CTC organizers\cite{Maska2023} using the normalized Acyclic Oriented Graph Matching (AOGM-D) measure\cite{Matula2015} for detection ($DET$), the Jaccard similarity index for segmentation ($SEG$) and the overall performance $OP_{CSB} = 0.5(DET+SEG)$. The AttUnet(XAI) model was quantitatively evaluated by the CTC organizers on the hidden test datasets. In the $OP_{CSB}$ metric, its performance peaked at 95\% for PhC-C2DH-U373, with a variation between 84.3\% and 95\% across the datasets. Segmentation accuracy, represented by the $SEG$ metric, ranged from 72.2\% to 91.7\%, and detection accuracy ($DET$) spanned from 88.5\% to 98.3\%. These results, are provided by the organizers of the challenge\cite{Maska2023}, demonstrate the model's consistent performance in key areas of segmentation and detection, underscoring its applicability in the field of automated cellular imagery analysis (refer to Tab.\ref{tab:TESTCTCseg}).

Despite the superior performance of top-performing approaches in CTC, such as CALT-US\cite{CALT-US}, ND-US\cite{ND-US}, and UNSW-AU (NAS~\cite{NAS}), it is important to note their limited availability for non-commercial, research purposes. This unavailability underscores the significance of our model, which, although not outperforming the leading methods, is accessible for research and potentially benefits from methodological enhancements tailored to specific datasets. Enhancements could include introducing a third class for cell borders, dataset-specific hyper-parameter tuning (nnU-Net~\cite{nnUNET}), advanced data augmentation techniques (random sequence reversals, random affine, and elastic transformations by NAS~\cite{NAS}), and incorporating manual annotation (BGU-IL~\cite{Maska2023}). Moreover, our approach, like most in the CTC, utilizes U-Net architectures, yet our model is notably more efficient, being significantly lighter with a file size of only 28.2MB. This is 11 times smaller compared to the nnU-Net, and at least 5 times smaller than BGU-IL's model for 2D datasets, demonstrating our model's efficiency.

The lightweight nature of our model, positions the Phagostat method as a baseline in automated cellular imagery analysis, particularly when evaluating performance relative to model size. By leveraging the AttUNet(XAI) model's compact computational footprint, Phagostat delivers rapid and precise cell segmentation analyses. This efficiency underscores its suitability for intensive workloads and environments constrained by hardware resources, highlighting its broad potential applicability in the field.\newline

\noindent\textit{Aggregates quantification:}
In order to accurately detect and quantify aggregates in time-lapse videos, we developed a specialized module that processes data through a series of steps. In summary, these steps include: (i) applying a threshold-based segmentation tailored for fluorescent aggregates; (ii) labeling the binary mask to separate individual aggregates; (iii) computing and saving the count, surface area, and coordinates of aggregates; and (iv) quantifying phagocytized aggregates based on changes in surface area and coordinates. Upon testing, we determined that utilizing the change in surface area of aggregates effectively captures two critical aspects: the surface area reduction when a cell begins internalizing an aggregate in a bite-by-bite manner, and the change in coordinates when the aggregate becomes small enough to be moved and internalized by the cell. By combining these two criteria, we can appropriately quantify phagocytosis, as when a given aggregate is fully internalized by a cell, it is considered to be phagocytized.

In our study, we utilized fluorescent aggregates, where high pixel intensities indicate the presence of an aggregate and low pixel intensities signify its absence. Consequently, a simple threshold was sufficient to accurately segment the aggregates. Furthermore, our experimental setup ensures that aggregates are initially fixed in the presence of cells. This implies that if the aggregates move, it is due to cellular activity. 

To delve deeper into the biological context, our imaging captures the dynamic interaction between microglia and fluorescent aggregates. The aggregates stand out with pronounced fluorescence against the background. During internalization, aggregates are partially obscured by the cell plasma membrane, causing transient moving white shadows due to cell movement. As microglia decompose the aggregates, the fluorescence diminishes, and the background darkens, ultimately turning pitch black when fluorescence stops. Our analysis technique distinguishes the genuine fluorescence signal of both stationary and moving aggregates from background noise through particle movement detection. This allows for consistent and precise identification and segmentation of non-internalized aggregates.

The reliability of this method hinges on two key assumptions: that the aggregates are fluorescent and initially stationary in the presence of cells. Any movement of aggregates, therefore, can be attributed to cellular activity. This is monitored through changes in the aggregate centroid from its initial position, a technique that might falter if the aggregates were not fixed or if the fluorescence assumption did not hold.

\subsubsection{Spatiotemporal superposition/analysis of aggregates and cells for phagocytosis quantification} 
To summarize the various modules of the presented pipeline, 'PhagoStat', which is dedicated to phagocytosis quantification (refer to Fig.\ref{fig1}): initially, computational resources are automatically allocated in the case of computation on a cluster (see Fig.\ref{fig2}.b); otherwise, a local machine is employed. The raw data is loaded, normalized, and separated into two channels for cells and aggregates (see Fig.\ref{fig2}.a). The aggregate data is utilized to correct any detected shift and remove blurry frames from the stack, if present (see Fig.\ref{fig3}.b). Subsequently, the aggregates are segmented, and their morphological features (i.e., area, count) are quantified (see Fig.\ref{fig3}.c). The cell data is processed through the scene instance-level cell segmentation module (see Fig.\ref{fig5}) to extract cell area and coordinates. These features are then used to track cells and estimate cell speed and total movement\cite{btrack1,btrack2}. Additionally, heat map visualizations and traceable time coherence data are provided (see Fig.~\ref{fig4}.c), resulting in a fully interpretable pipeline.

All the results are compiled and organized by condition to generate a comprehensive statistical report. It should be noted that, currently, the statistical reporting is limited to only two conditions. Therefore, in cases where more than two conditions are being used, the reporting framework would need to be adapted accordingly to accommodate the additional conditions.

To provide a sense of the efficiency of our pipeline, it takes approximately 20-30 minutes to process raw data and generate the final statistical report. This processing time applies to 20 scenes, each lasting for 7 hours, with aggregate and cell frame pairs captured every 2 minutes. The pipeline's performance was evaluated using a high-performance computing cluster equipped with only Central Processing Units (CPUs). It is expected that the processing time could be further reduced if Graphics Processing Units (GPUs) were utilized, highlighting the pipeline's potential for rapid and effective analysis.

\subsection{Microglial cells phagocytosis use case}
\label{microglial-use-case}
The dual role of microglial phagocytosis, encompassing aggregate clearance and the abnormal phagocytosis of live neurons and synapses, has been a topic of investigation for several years in various neurodegenerative diseases, such as Alzheimer's and Parkinson's diseases~\cite{A,B,C,D,E}. In order to establish our assay, we focused on the case of FTD mutations, where the two most frequently mutated genes in familial forms of the disease -- \textit{C9ORF72} and \textit{GRN} -- are highly expressed in microglia and are believed to regulate microglial functions, particularly phagocytosis~\cite{H,I,J}. Given that TDP-43 aggregates accumulate in the degenerating neurons of patients~\cite{F,G}, aggregate clearance is anticipated to yield positive therapeutic outcomes. Conversely, excessive synaptic pruning or phagocytosis of live neurons could exacerbate degeneration. This has been demonstrated in the case of C9ORF72, where mutations involved in neurodegenerative diseases can influence both opposing effects~\cite{I}.

Moreover, numerous simple phagocytosis assays exist, utilizing latex beads or pH-sensitive fluorescent particles. Nonetheless, the development of sensitive assays capable of analyzing phagocytosis using physiological targets, such as protein aggregates and functional neuronal networks, and breaking down microglial phagocytic behavior is of major interest. This approach is crucial for the development and screening of therapeutic compounds aimed at targeting abnormal phagocytic activities.

In an effort to estimate the on-target phagocytic activity of WT and FTD-mutant microglial cells, we analyzed the amount of TDP-43 aggregates internalized per cell (Fig.~\ref{fig6}.g: aggregates ratio = area eaten/cell count). Interestingly, FTD mutant cells seem to display increased phagocytosis of TDP-43 aggregates, with FTD mutants being approximately 70\% more aggressive than their WT counterparts. We ensured that the number of cells in the assay (Fig.\ref{fig6}.h: cell count) was similar, yet we observed that the size of the cells was significantly larger in FTD-mutant microglial cells (Fig.\ref{fig6}.i: mean cell area). Notably, FTD mutants were approximately 30\% larger than WT cells, a finding that, to our knowledge, has not been proven before. As a result, we measured the amount of TDP-43 internalized per cell surface unit (Fig.~\ref{fig6}.j: aggregates ratio = area eaten/cell area) and found that a portion of the increased phagocytic activity of FTD-mutant microglial cells could be explained by the increased spreading of the cells. Moreover, no global modification of the total quantity of movements or the speed of the movements of the cells was observed in this case (refer to extended data~\ref{extended}: Fig.~\ref{sup_fig_2} and Fig.\ref{sup_fig_4}).

\subsection{Phagocytosis dataset for microglial cell}
In this study, we have chosen to analyze the phagocytosis of protein aggregates by microglia in the context of frontotemporal dementia (FTD). FTD is a neurodegenerative disease in which mutations in genes regulating microglial functions (such as \textit{C9ORF72} and \textit{GRN}) correlate with a specific type of aggregates composed of the TDP-43 protein~\cite{K, F, G}. Since the distinction between \textit{C9ORF72} or \textit{GRN} is not pertinent to the scope of this work, we will refer to FTD mutants for simplicity.

The data acquired for this study consists of acquisitions from wild-type (WT, n=5) and frontotemporal dementia (FTD, n=5) microglial cells during the phagocytosis process. Each acquisition captured 20 distinct scenes, encompassing seven hours of time-lapse video microscopy (recorded at one frame every two minutes). Cell and aggregate images were collected in two separate channels.

Our dedicated team of biologists has meticulously generated a comprehensive dataset, which has been subjected to rigorous validation by the laboratory's ethical committee. This dataset encompasses 36496 normalized cell and 36496 aggregate images, accompanied by 1306131 individual instance masks for cells and 1553036 for aggregates, 36496 registered aggregates and data of the intermediate steps in tabular format all generated using the PhagoStat algorithm. To ensure the highest quality of data, we have applied data quality correction techniques.

The dataset provides an extensive array of biological features, such as area, position, and speed, which are presented in a frame-by-frame manner to allow for in-depth analysis.

Moreover, to enhance the dataset's value, we have incorporated 226 manually annotated images containing 6,430 individual cell masks (seed dataset). These images were randomly selected from different conditions, including wild-type (WT) and frontotemporal dementia (FTD), and were obtained from a diverse range of randomly chosen scenes. To ensure precise annotations, we started with a polygon-based method using (labelme: \url{https://github.com/wkentaro/labelme}), where our team of expert biologists annotated 61 images. Despite their diligence, the process was inherently slow and prone to inconsistencies given the irregular shapes of the cells. To counteract these limitations, we developed "Point2Cell," a GUI annotation tool (source code at \url{https://github.com/ounissimehdi/Point2Cell}), which markedly improved annotation efficiency. It boosted the Dice score to 94.97\% (compared to 91.3\%) for precision and significantly reduced the annotation time (from 96.1 sec to 14.6 sec per image). This leap in performance led us to use Point2Cell for annotating the subsequent 165 images (4,694 individual cell masks), which we then designated as the test set to objectively evaluate our models. Meanwhile, the initial 61 images (1,736 individual cell masks) annotated via "labelme" were retained for training and validation purposes, ensuring our models were well-tuned before being tested.

The result is a robust dataset comprising a total of 235,288 files (94GB) of '2D + time' aggregate and cell in mono-layer configuration each, offering researchers a valuable resource for investigating various cellular and aggregate properties in their studies.

Firstly, our dataset can serve as a benchmark (available at \url{https://zenodo.org/records/10803492}), providing a reliable and comprehensive reference point for researchers to gauge the effectiveness of their algorithms or techniques vis-à-vis PhagoStat. This meticulously generated data, which encompasses both WT and FTD mutant microglial cells during the phagocytosis process, laying a robust foundation for comparative analyses.

Secondly, the depth and breadth of our dataset — detailing aspects such as area, position, and speed on a frame-by-frame basis — acts as a definitive proof of concept. It showcases PhagoStat's capability in analyzing the phagocytosis of protein aggregates by microglia, particularly in the context of diseases like FTD.

Lastly, for those aiming to broaden PhagoStat's utility across various data, our dataset can significantly contribute to the pre-training stage. Using our dataset as a foundation for model pre-training, may speed up convergence and improve generalization capabilities. To streamline this process, we have applied AttUNet(XAI) and UNet(XAI) to datasets (WT-1 to WT-3 and FTD-1 to FTD-3) for training, and used (WT-4 and FTD-4) for validation. Upon evaluation with (WT-5 and FTD-5), both models attained an exceptional Dice score of 97.88\%.

\subsection{Enhancing interpretability in deep learning}\label{sec:interpretability-heatmap}

\subsubsection{Model sensitivity analysis methodology for smart annotation in cell image segmentation}

The results of our study emphasize the importance of developing visualization tools for direct model ablations, specifically focusing on the effects of input modifications on model parameters during training and their subsequent impact on outputs during the testing phase. This process, typically time-intensive due to prolonged training, validation, and visualization phases, presents significant challenges in practical applications. However, as outlined in the methods section: 'Automated DL model optimization using feature maps' and as illustrated in Fig.~\ref{sup_fig_features}, our approach effectively addresses these challenges. By reducing model size, we not only optimize the training time but also minimize the risk of over-fitting, a crucial factor in training models with limited input images to ensure the validity of our conclusions.

In tasks like biological study-related segmentation, choosing the \textit{right} images for annotation is a critical, and often a random process, limited by the available resources. Our methodology aids in selecting the most beneficial types of images for model training. We tested this capability by assessing the impact of different quantities of annotated cell masks on the training performance, focusing on cell count variation. Our experimental setup uses three images per cell count from our microglial dataset (randomly selected from distinct acquisitions to avoid data leakage). We explored 13 conditions with 14 to 40 cells per image, training distinct models (from scratch) for each condition using the Attunet(XAI). This process required less than 30 minutes to train 13 model using a standard GPU, so demonstrating the effectiveness of our approach in rapidly assessing model performance across varying conditions.

Our findings provide valuable guidance for annotators (refer to Fig.\ref{sensitivity_dl}.left), suggesting the prioritization of certain images that significantly impact model performance. For instance, the top-5 performing models indicate that images with cell counts between 28 and 38 and a foreground ratio of 31\% to 47\% are optimal for model training. To confirm these guidelines, we conducted extended testing with 2,300 images (refer to Fig.\ref{sensitivity_dl}.right), which reinforced the initial recommendations derived from a much smaller sample. This extensive testing underlines the model's sensitivity to specific parameters (like cell count) and showcases the  efficiency of our method in evaluating the behavior of the model with minimal annotation and training requirements.

Overall, our study presents a streamlined, efficient approach for model assessment and annotation strategy in biological segmentation tasks. By requiring minimal resources and time, our methodology not only aids biologists in smart annotation but also establishes a practical framework for exploring model behavior in controlled settings. To further assist in the application of our findings, we have made the source code available in notebooks on Github, completed with visualization tools for immediate analysis, thus simplifying the integration process for various conditions.

\subsubsection{Heatmap visualizations for XAI}
In the domain of neuroscience, heatmaps provide an intuitive representation of feature intensity and distribution across spatial dimensions, closely mirroring the conceptual framework of neural activity and feature relevance. These visual tools offer a stark visual contrast, facilitating the identification of patterns and significantly aiding in pattern recognition at a glance.

Moreover, the prevalent use of heatmaps in bio-informatics and neuroimaging means that they are well-established within the scientific community, likely reducing the cognitive load for domain experts when interpreting these visualizations. This familiarity is bolstered by the computational efficiency of heatmaps, which proves beneficial during the iterative development and inferential stages of model training.

While recent advances in deep learning visualization techniques—like class activation mapping (CAM) methods\cite{cam}, gradient-based approaches\cite{gradcam, gradcam++}, dimensionality reduction via t-SNE\cite{tsne}, and feature relevance assessments using SHAP values\cite{SHAP_1,SHAP_2,SHAP_3,SHAP_4}—have proven effective for classification tasks, when the spatial information is compromised during the fully connected layer transition. Our study, however, is directed toward dense binary segmentation tasks, meticulously preserving spatial detail at the pixel level. By maintaining spatial fidelity across network operations (including max pooling and up-sampling) our methodology not only preserves spatial integrity but also provides a practical and cost-effective solution for pixel-wise tasks.

The deployment of heatmaps in the training phase of our DL model offers a compelling narrative of the model's learning process, enhancing trust among biologists and neuroscientists. We have documented key stages in model training—starting from the initial discernment of cell textures to the sophisticated recognition of intracellular elements and individual cells, even under partial visibility (refer to Fig.\ref{DL_evo}). This progressive refinement, captured through heatmaps, provides biologists and neuroscientists with a tangible understanding of the model's feature learning, resonating with their empirical work and theoretical constructs.

Furthermore, heatmap visualizations act as a conduit, demystifying the intricacies of DL by providing concrete, visual progressions that domain experts can readily assimilate. These tools are indispensable not only in elucidating the rationale behind the model's decisions but also in steering further enhancements of the architecture, as elaborated in our methods section ‘Automated DL model optimization using feature maps’.

Through continuous dialogue with neuroscientists, informed by these visualizations, our methodology has been refined, emphasizing the practical significance of heatmaps in an interdisciplinary nexus. Although a formal user study to evaluate this approach is forthcoming, preliminary feedback suggests that this visualization technique significantly augments the interpretability and receptivity of DL models within the biologists and neuroscience communities. Future research endeavors will be directed towards quantifying the impact of these methodologies, with the goal of incorporating them into a framework that allows experts to provide real-time feedback. This interactive process is designed to refine the training protocol of the model and enhance its decision-making capability in real-time production, ensuring alignment with specialized expert knowledge.

\section*{Discussion}
Phagocytosis is a vital cellular process that serves as a primary defense mechanism against danger signals and pathogens, ensuring the proper functioning of the immune system. In the brain, microglial cells are the sole cell type capable of performing phagocytosis. The phagocytic activity of these cells has garnered increasing attention in the study of neurodegenerative diseases, as neuroinflammation appears to play a important role in the pathological processes. This role may involve the clearance of aggregate formations or the abnormal phagocytosis of live neurons and synapses by microglia. Furthermore, antibody-mediated clearance of aggregates by phagocytic microglia is currently being explored as a therapeutic strategy for Alzheimer's disease and other forms of dementia.

Quantifying the phagocytosis of amorphous and highly active unstained cells is a challenging task, which is essential for investigating a wide range of neurodegenerative diseases. This task often involves phase-contrast time-lapse video microscopy to observe rapid interactions between cells, making it difficult to distinguish the cells of interest from the background.

In response to these  challenges, PhagoStat offers a scalable, real-time analysis end-to-end framework adept at leveraging HPC clusters to process vast datasets swiftly, exemplified by its ability to handle 750 GB across ten CZI files (200 unique sequences of 7h) in just 97 minutes using only CPU power. Moreover, its adaptability to hardware advances, such as tripling data retrieval speeds when switching from HDD to SSD storage, promises continual performance enhancements. PhagoStat's design philosophy aligns with its principles, ensuring that data analysis keeps pace with data acquisition. It's capable of concurrently processing data as it's being recorded, which allows for the completion of data analysis from a 7-hour recording within just 20 minutes. This synchronization with the microscope's functionality means that analysis is ready immediately upon recording completion, thus streamlining the entire workflow. 

PhagoStat ensures transparency by making intermediate results accessible, fostering trust and comprehension among its users. Moreover, its design aligns with General Data Protection Regulation (GDPR) guidelines, emphasizing reproducibility—integral aspects that not only bolster scientific integrity but also expedite the translation of research into practice. This alignment with GDPR not only reinforces the safety and traceability of data but also support PhagoStat's position as an innovative tool designed for the future of biological data analysis.

By providing an interpretable and transparent pipeline, PhagoStat empowers experts and users, such as laboratory technicians, biologists, physicians, and supervisors, to gain a deeper understanding of the studied processes. Additionally, this approach opens up exciting possibilities for pipeline design optimization tailored to specific use cases. As modern technologies increasingly demand a lower carbon footprint, PhagoStat offers valuable insights into the design of individual modules, ensuring that the pipeline remains both efficient and environmentally responsible.

Effective data quality check capabilities are essential for adapting to real-world acquisition conditions in scientific research. One limitation of our current approach, however, is the need for optimization improvements in our registration method, the CECC, which takes longer than the widely-used SIFT method. It is crucial to further investigate the bias detected when using the SIFT method, as this may offer valuable insights for enhancing our pipeline.

Our analysis suggests that the bias observed in the SIFT method could be related to issues with landmark identification. We believe that this issue may arise from at least one of the following reasons: 1) an insufficient number of landmarks, as the signal is predominantly background, or 2) indistinguishable landmark features due to the geometric similarity of aggregates. Even though our CECC method appears to be immune to these issues, it comes at the expense of speed. It is essential to further investigate these potential causes to ensure the robustness and accuracy of our pipeline, especially if SIFT is considered as an alternative to CECC for faster results

Transitioning from a DL-only to an IDL approach has yielded a substantial decrease in model size by a factor of seven, across the board. This shift is underpinned by the strategic implementation of an automated evaluation system for feature map integrity, utilizing a precise metric that not only streamlines the selection of optimized models but also conserves valuable time for both experts and computational resources. The result is a more efficient methodology that broadens the usability of these models. Indeed, they demand lower hardware specifications while maintaining high-performance level of their larger counterparts.

Despite the promising properties of the TTCM described in this work, TTCM relies heavily on the UNet predictions at different time points. This will yield good results only if UNet-like models are well-performing on the test distribution. One possible direction for improvement could involve the full integration of TTCM into the UNet during training, supplemented by a time coherence loss function.

In this study, we employed FTD-mutants and WT cells as a use case to demonstrate the utility of our pipeline. For the sake of simplicity, we grouped GRN and C9ORF72 mutations together as FTD-mutants. It is important to note that there may be biological differences between GRN and C9ORF72 mutations, which are beyond the scope of this work. The investigation of these differences should be addressed in future studies, in order to enhance our understanding of the specific characteristics and implications associated with each mutation type. By elucidating these distinctions, researchers can potentially develop more targeted and effective therapeutic strategies for neurodegenerative diseases involving these mutations.

In this study, our main contribution is the study, the development and the public release of the PhagoStat pipeline, an unprecedented end-to-end solution for data-sequence management, video analysis, noise handling, quantitative analysis, and statistical reporting. Additionally, we have applied the PhagoStat pipeline to microglial cells, revealing novel insights into FTD, notably the increased size and activity in mutant cells compared to wild types. This finding offers significant contributions to neurodegenerative disease research. Complementing this contributions, we are releasing an extensive, first-of-its-kind dataset on microglial cell phagocytosis 2D+time, inclusive of various cell and aggregate images with their masks, to further support and inspire community research and development.

As we pivot towards the future, we are captivated by the prospect of delving into 3D spatio-temporal analysis. This is not merely a shift in dimensionality but a stride towards an authentic in vivo understanding of cellular behaviors—akin to observing life through a more revealing lens.

The transition into 3D analysis brings forth formidable challenges in computational processing and data management. These challenges herald a pioneering phase in methodological computer vision and interpretable AI, essential for methodological advancements. Such innovations are crucial for neurodegenerative disease models, such as brain organoids, where sophisticated imaging and analytical methods are urgently needed.

Our commitment to leading this innovative field is underscored by integrating advanced imaging techniques into our ongoing research. Our aim is to establish a strong, reliable framework within the discipline that significantly improves the analysis of complex biological data. This, in turn, will directly benefit researchers specializing in microglial cell phagocytosis, offering them unprecedented insights and tools. This endeavor extends beyond the advancement of our own methods; it enriches the scientific community's resources, propelling our methodology forward. By sharing our approaches and data, we contribute to a shared foundation upon which the scientific community can build, fostering a more profound understanding and facilitating further breakthroughs in this vital area of study.

\section*{Methods}

\begin{figure}[H]
    \centering
    \includegraphics[width=\textwidth]{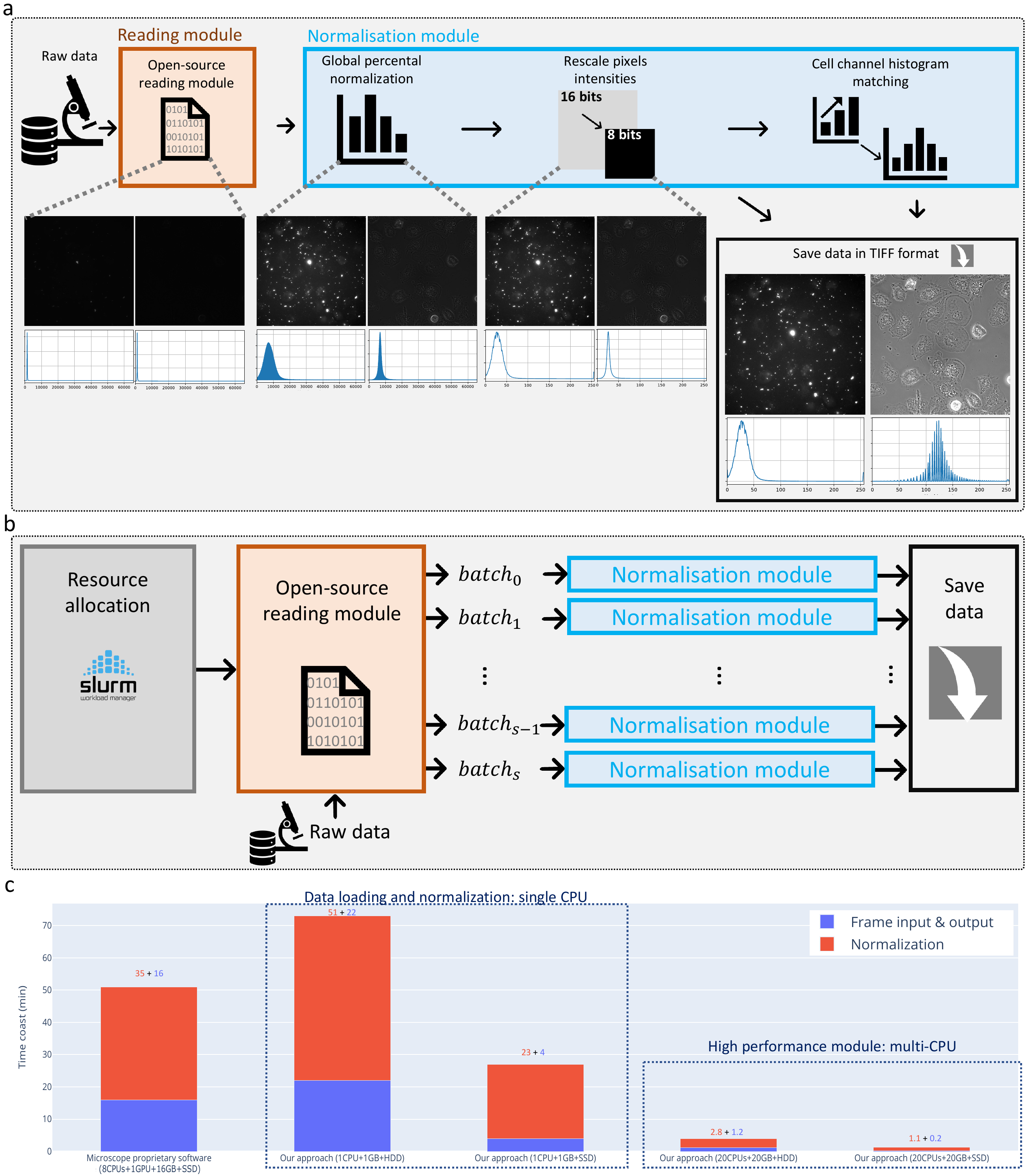}
    \caption{\textbf{Efficient data loading and normalization pipeline}. \textbf{(a)} Detailed steps of the data loading and normalization module where: the two channels (aggregates and cells) are extracted directly from the microscope raw data, then it applies the local and global normalization to standardize the data. \textbf{(b)} High performance computing (HPC) cluster compatible scheme that scales to big datasets. \textbf{(c)} Quantitative comparison of our single-CPU/multi-CPU method and the GPU-accelerated Carl Zeiss ZEN software when processing 76GB CZI file (raw data). To facilitate direct comparison, 'Frame input \& output' times are the combination of "read and write" times for all systems. As an insight, our method's time allocation for SSDs (25\% reading and 75\% saving) and HDDs (76.6\% reading and 23.3\% saving).}
    \label{fig2}
\end{figure}

\begin{figure}[H]
    \centering
    \includegraphics[width=0.94\textwidth]{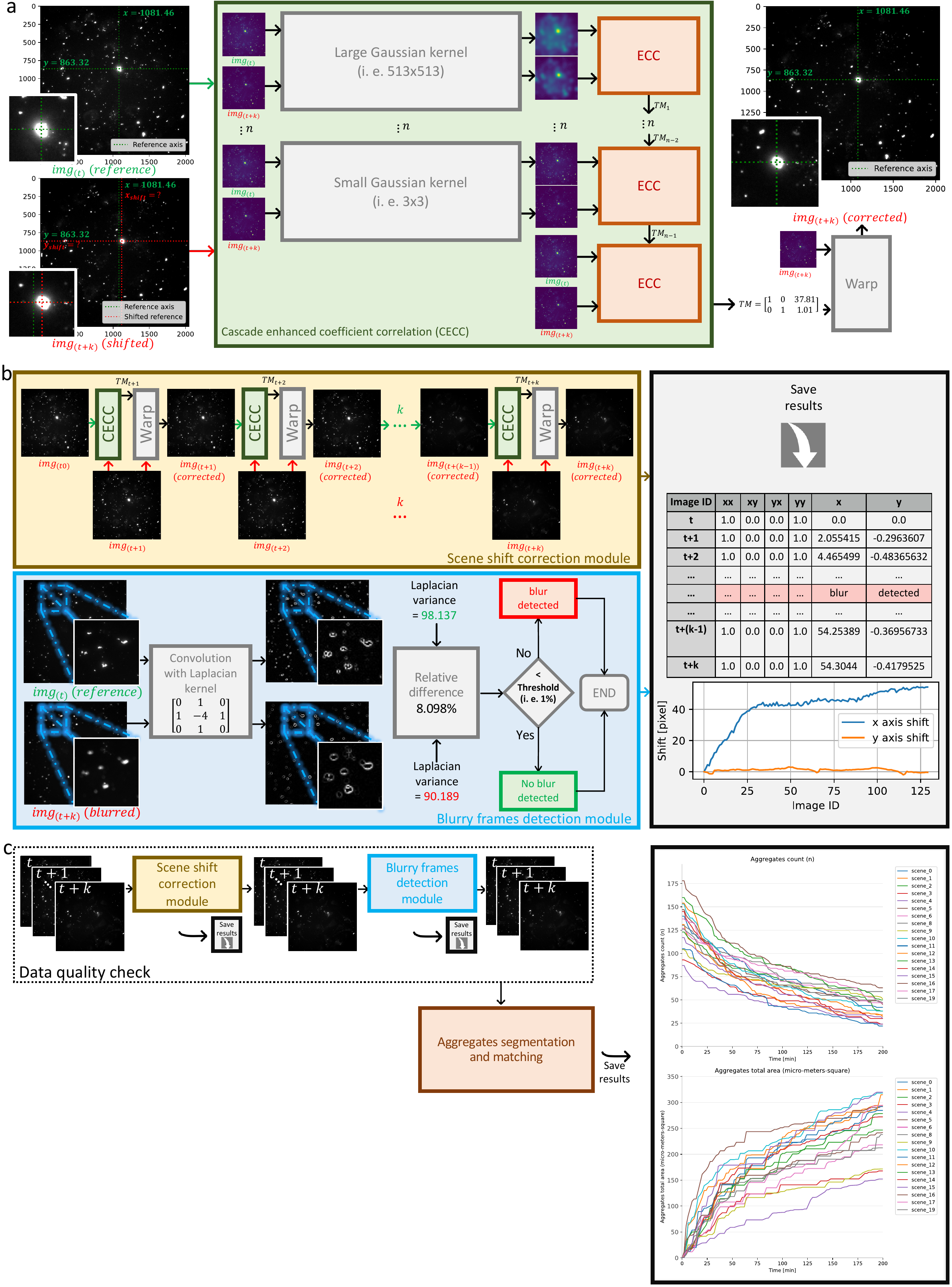}
    \caption{\textbf{Detailed data quality workflow. (a)} Detailed CECC registration approach. \textbf{(b)} Detailed data quality check modules: (i) CECC based scene shift correction module, (ii) blurry frames detection module, and (iii) saving registration information and the rejected blurry frames. \textbf{(c)} Overview of the aggregates quantification workflow: data quality check + segmentation and matching.}
    \label{fig3}
\end{figure}

\noindent\textbf{Data efficient loading and normalization}\label{norm_meth}.
Most of the proprietary microscope software (biologists friendly) are working exclusively on Windows. Therefore, data preparation required specific steps: (i) raw data was first transferred from the microscope machine to a Windows machine; (ii) the acquisition was converted into tagged image file format (TIFF) frames; (iii) the frames were arranged to be compatible with the computational pipeline and iv) the resulting frames were transferred to a high performance computing (HPC) cluster for processing. A notable challenge in this process was the lack of transparency we found when running the software's preprocessing (black-box) steps (i.e., normalization, the conversion of the raw data 16bit to TIFF 8bit frames). Besides, the rich and user-friendly visualization interface, such packages turned out to be overly opaque, inefficient, resource-intensive and time-consuming for analyzing big data. To overcome these drawbacks, we have adapted the 'aicspylibczi'\cite{aicspylibczi} to develop a flexible, robust, and open-source module capable of reading and converting proprietary raw data formats into universal image formats. Our module was tested on converting Carl Zeiss Image (CZI) files to TIFF format, and it is compatible with Windows, macOS, and Unix. Additionally, the module can be easily adapted to take advantage of HPC clusters and parallelization schemes (refer to Fig.\ref{fig2}.a and Fig.\ref{fig2}.b). 

The 'aicspylibczi'\cite{aicspylibczi} Python package was used and extended to read the 'CZI' raw data file using delayed reading. This approach allowed us to read cell and aggregate channels image-wise without loading all the sequences to the RAM. Images were in 16bit representation. Frame reading time can be accelerated by coupling the Python package with multiple CPUs for parallel processing. This involves several scenes being simultaneously read, with each scene being allocated to its own CPU. This deviates from sequential processing, in which scenes are placed in a queue and read sequentially; (i) in the local machine, our package can use multi-CPUs for parallelism, or (ii) in a HPC clusters that use simple Linux utility for resource Management (SLURM), where our package launches an array of jobs (i.e., attributing to each scene a job ID) on the same node or different nodes.

While using the same parallelism scheme, global percentile normalization is used to re-scale the image pixels' intensities of the whole sequence; 0.5\% and 99.5\% percentiles for aggregates; 0\%, 100\% for cells. Aggregate and cell images were re-scaled from 16bit to 8bit using 'img\_as\_ubyte' function from the 'scikit-image'\cite{van2014scikit} Python package. Histogram matching with a normal pixel distribution as reference is used on all cell images, and we apply it using the 'match\_histograms' function from the 'scikit-image' Python package. Finally, if needed, images were resized from $2048\times2048$ to $1024\times1024$, then saved in 'TIF' format using the 'PIL' Python package.\newline

\noindent\textit{Isolating aggregate and cell signals for precise quantification and segmentation.} In typical microscopy practices, biologists rely on default software provided with the microscope, which often merges the signals from cell instances and aggregates. This fusion allows for visual interpretation, although it may not always suit quantitative analyses or automated processing.

Our approach diverges by treating these channels distinctly to facilitate precise quantification. Specifically, the aggregate channel, which we use to identify clusters of particles, is marked by a unique fluorescent tag (for instance, a red chromatin signature). This enables us to isolate these aggregates onto a separate grayscale image layer using fluorescence microscopy.

Simultaneously, the non-fluorescent signal, corresponding to individual cell instances, is captured on a different grayscale layer. This separation is crucial because it allows us to apply specialized image processing techniques to each channel independently, enhancing the accuracy of cell instance segmentation and aggregate quantification.

During data loading in our system, we maintain this separation. Each channel is loaded individually, thereby preserving the integrity of the information they contain. Consequently, this segregation of data not only simplifies the subsequent image analysis but also ensures that any computational models or algorithms applied later can be fine-tuned to the characteristics of each channel without cross-contamination of signals.

\noindent\textit{8-bit conversion for performance, transparency, and storage.} We evaluated the impact of image bit-depth on the performance of convolutional neural networks in segmentation tasks. The decision to employ an 8-bit conversion of images was informed by a comprehensive ablation study. This study involved training two identical Unet(XAI) networks: one with 16-bit image inputs and the other with 8-bit image inputs. The findings indicated a marginal enhancement in segmentation performance for the 8-bit model, with mean test losses registering at 0.0802 compared to 0.0827 for the 16-bit counterpart. These results suggested that the lower bit-depth conversion did not hinder, and may in fact have slightly improved, model efficiency for our microglial-dataset.

We used the Fluo-N2DL-HeLa dataset, publicly available from the CTC. Notably, the images in this dataset are inherently 16-bit, and correspondingly, the annotations were also done in 16-bit. Our approach involved training two UNet(XAI) models, one with 16-bit images and the other with 8-bit images. Both models were trained using identical random seeds. We adopted a specific training methodology: we utilized the first sequence of the dataset for training and validation, while the second sequence was reserved for testing. The performance of each model was quantitatively assessed. For the 16 bit model: test loss = 0.1464 and Dice = 0.9642, in contrast with the 8 bit model: test loss = 0.1514 and Dice = 0.9621

We analyzed the loss of information when images are converted from 16-bit to 8-bit format. The process began with re-scaling the intensity of 16-bit images to utilize the full 16-bit range. Following this, we converted these images into 8-bit format. Besides, we divide 16-bit images by $2^{16}$ and 8-bit images by $2^8$ for a direct comparison (images between 0 and 1). Then, we measured the Mean Squared Error (MSE), Structural Similarity Index (SSIM), and Peak Signal-to-Noise Ratio (PSNR) between the re-scaled 16-bit and the converted 8-bit images. This methodology provided a thorough assessment of how the image quality and information fidelity are impacted by the conversion process. The analysis yielded the following results: $MSE = 1.3698e-06 \pm 5.5903e-08$; $PSNR = 58.6368 \pm 0.1747$; $SSIM 0.9996 \pm 2.2954e-05$.

The results across the test loss and Dice show negligible loss (0.005 difference in the test loss and 0,0021 difference in the Dice score), also, across MSE, PSNR and SSIM metrics show negligible loss, thereby providing evidence that the conversion between 16-bit and 8-bit imaging does not significantly impact the model training (at least on the microglial-dataset and the Fluo-N2DL-HeLa dataset).

Furthermore, in collaboration with domain experts in biology and neuroscience, we recognized the necessity for a transparent analytical pipeline. Our collaborators expressed the need for visibility into the intermediate processing stages of CZI file handling to establish trust and ensure traceability of results. To address this request, we incorporated intermediate outputs in universally accessible formats such as GIF, TIFF, and PNG, thus enhancing the interpretability of our pipeline.

Storage optimization also played a pivotal role in our methodology. The original microscope images were captured at a resolution of 2048x2048 pixels in 16-bit format. Given the substantial data storage requirements, particularly when retaining all intermediate outputs, we implemented a strategy to resize images to 1024x1024 pixels in 8-bit format. This approach reduced the storage per frame from approximately 4.2MB to 0.6MB, achieving a nearly seven-fold decrease in data size. This significant reduction facilitated easier data sharing and handling within the research community, enabling peers to download and utilize our dataset more efficiently.

In summary, our methodological adaptations, particularly the conversion to 8-bit image processing, have resulted in an efficient, transparent, and storage-optimized pipeline without sacrificing the integrity and performance of our deep learning models.\newline

\noindent\textit{Quantitative performance evaluation of the readout module}. The proprietary Carl Zeiss Microscopy ZEN light v3.3.89.00 software was used as a reference for data loading and saving. This proprietary software was evaluated using a Windows 11 machine with 8 cores i7 9700K CPU, 16GB RAM, Nvidia RTX2080 GPU and Samsung 970 PRO SSD; all drivers were up-to-date. CUDA acceleration was enabled from the software configuration panel; all parameters were left at their default values, and no tasks ran in the background before/during the benchmark.

Our approach uses only open-source Python packages, as described and cited before. Ubuntu 20.04LTS was used with: 8 cores, i7 9700K CPU, 16GB RAM, HDD or SSD. For single-CPU tests, the hardware used was limited to 1 CPU using 'taskset'. The test was monitored, and during the test, RAM usage did not exceed 1GB (while using an HDD or SSD). For the multi-CPU test, SLURM was used to process 20 job arrays. Each one uses 1 core Xeon Gold 6126 CPU and 1GB RAM (while using HDD or SSD storage node).

All data transfers (single raw data file or frames) were performed using a 1GB/s Ethernet port with 'FileZilla' v3.46.3. SFTP transfer protocol was used while directly connected to the internal institute network (no VPN used); maximum simultaneous transfers were set to 10 files (FileZilla's upper bond).\newline

\noindent\textit{Technical aspects of our data normalization strategy for large-scale video-microscopy datasets.} We initially considered the straightforward approach of loading complete CZI files into a numpy array for processing. This method is feasible for sequences with limited data or shorter duration, given a conventional computational setup with 16/32GB of RAM. However, this became impractical with our datasets, where individual CZI files were approximately 76GB each, comprising 20 sequences captured over 7 hours with a two-minute frame interval across two imaging channels. Standard computing resources were quickly overwhelmed by these files, as evidenced by system performance when RAM capacity was exceeded (refer to Fig.~\ref{fig2}.c), leading to the use of SWAP space which is much slower than RAM.

In the realm of HPC environments, even though they offer a more robust infrastructure, the extensive size of our datasets still resulted in considerable processing delays and resource contention. This was due to the immense memory requirements (80GB of RAM per CZI file) which induced periods of computational idleness within the HPC's dynamic job scheduling system.

To circumvent these bottlenecks, we adopted a delayed data reading methodology. This technique does not enhance the speed of normalization computation directly but rather optimizes memory consumption. By strategically fetching only the necessary data segment for processing—such as a single frame from a sequence—we were able to initiate 20 Slurm jobs (one job per sequence) in parallel, each consuming only 1GB of RAM as opposed to the full 80GB that the entire file would require. This also enabled selective access to data, permitting us to isolate specific frames from the imaging channel for detailed morphological analysis.

Technical differentiation between "global" and "local" normalization is crucial in our work. "Global" normalization refers to the standardization of intensities across the entirety of each sequence, whereas "local" normalization pertains to adjustments made on a per-frame basis within a sequence. For global normalization of the aggregate channel, we utilize the first frame's 0.5\% and 99.5\% intensity percentiles as a baseline (reducing outlier intensities and improving contrast), since all aggregates are present at the beginning of the experiments (non phagocyted yet). Subsequent frames are normalized against this reference, ensuring consistent visualization of morphological features. For the cell channel we use intensities (0\% to 100\% percentiles) from the first frame to adjust each frame thereafter. Following background noise reduction, we observed a Gaussian distribution in pixel intensities, prompting the adoption of a normal distribution model for local histogram normalization. This method effectively enhances contrast and feature prominence without altering the intrinsic cellular characteristics.

For the aggregate channel, only the percentile values of the first frame are loaded for global normalization, with the rest of the sequence can be processed in parallel. Similarly, for the cell channel, we load the first frame’s percentile values and the parameters for the Gaussian distribution, with the rest of the sequence can be processed in parallel.

To conduct an ablation study on the impact of histogram normalization, two UNet models were trained using the same seed dataset for training and validation. When evaluated on the test set, the model utilizing histogram normalization demonstrated up to a 10\% improvement in the Dice coefficient, indicating its effectiveness.

In summary, our method of delayed reading, coupled with parallel processing, has been meticulously designed to tackle the challenges presented by extensive datasets and the practical limitations of available computational resources. This detailed explanation should provide a comprehensive understanding of our data normalization methods and the technical reasoning behind our approach.\newline

\noindent\textbf{Frame registration and correction}. We applied SIFT~\cite{SIFT} algorithm to two frames affected by the shift problem. First, we identified the main points of interest and cross-referenced them with the next frame. Then, the outliers were discarded to estimate the transformation matrix, thereby canceling the shift. However, according to our performance evaluation, SIFT was sufficient to correct the shift (see Fig.~\ref{fig6}.a also in extended data \ref{extended} Table.\ref{tab:reg}). Finally, it obtained an average error of $0.0153\pm0.0609px$ along the x-axis, while $0.0228\pm0.1221px$ along the y-axis. Thus, SIFT proved to be directionally biased by the shift. SIFT was tested using the 'SIFT\_create' function from the 'OpenCV'\cite{bradski2000opencv} library to compute key points and their source and target image descriptors. The Euclidean distance (default sift error=0.7) matches points between the two key-points descriptors. The random sample consensus (RANSAC) algorithm eliminates outliers (the 'RANSACRegressor' function from the 'scikit-learn' library). The matched points are used to find the transformation matrix.

Our recursive scheme (in our CECC module) contemplates the intricacies and challenges of this registration task. Initially, we employed a relatively large Gaussian kernel of size 513x513 ('getGuassianKernel' function from OpenCV), where \(\sigma~=~0.3~\cdot~((kernel\_size-1)\cdot0.5-1)+0.8)\), which corresponds to a \(\sigma\) value of approximately 77.3. This choice was made to retain the essential details of the aggregates while diminishing noise. Using this, we computed an initial transformation matrix \(TM_{0}\) to adjust for the offset between successive frames. Subsequently, we transitioned to a smaller Gaussian kernel, specifically 257x257, equivalent to a \(\sigma\) of roughly 38.9. This finer kernel resolution introduced more granularity in the aggregate details. With the previously estimated \(TM_{0}\) as a starting point, we initialized a second registration process, leading to \(TM_{1}\). This new transformation matrix proved to be more adept at countering the acquisition shift. We iteratively proceeded with this method, progressively reducing the kernel size during each step until we reached a point where no kernel was necessary. The final transformation matrix, \(TM_{N}\), was determined using ECCM directly on the untouched frames, taking \(TM_{N-1}\) as its initialization.

To implement the CECC approach, we used the ECC implemented in the 'OpenCV' v4.5.1 Python library named 'findTransformECC'. Each cascade used a different Gaussian kernel, with 1000 max iterations or $10^{-4}$ error as a termination criterion for finding the corresponding transformation matrix. The last cascade computed the effective transformation matrix. 

When the transformation matrix is estimated, the 'warpAffine' function (from the 'OpenCV' library) is used to register the image by the computed transformation matrix.

In order to validate our registration approach, 1000 'x' and 'y' shifts were randomly generated and then saved between -400 pixels and 400 pixels (x and y shifts are independent). For each test, we loaded the reference image (containing aggregates) and the same random shifts, in the exact same order. We created a shifted version of the reference image using 'warpAffine', with the loaded shifted 'x' and 'y'. Both images (reference and shifted) are $2048\times2048$ gray-scale. Each test was submitted as a job via SLURM to a computational cluster. For CECC, we used 4 cores Xeon Gold 6126 CPU, 1GB RAM and for SIFT 4 cores Xeon Gold 6126 CPU, 2GB RAM (1GB RAM for SIFT is not sufficient). The execution time is computed and reported for each registered image.

For blurry frame detection module, we computed the Laplacian of two images: \textit{image(t)} and \textit{image(t+k)}, where k is the step between two images (i.e., k=1 means comparing two consecutive images). Then, the module evaluate the variance of the resulting images. Images with no blur give high variance values, and images with blur give low variance values. This mechanism effectively detects sudden drops in Laplacian variance values (using the relative difference compared to a given threshold), thus detecting blurry and unusable frames (i.e. dropped from the stack). 

In order to compute the Laplacian image, we used the 'Laplacian' function from the 'OpenCV' Python library in 64 float representation. Variance is then computed on the resulting image. Every two consecutive frames, the relative difference is computed, and the blurriness is detected if:
\begin{equation}
\lvert1-\frac{\sigma^2(\nabla^{2}_{5} f_{t+1})}{\sigma^2(\nabla^{2}_{5} f_{t})}\rvert > \varepsilon_{blur},~where~\varepsilon_{blur}\in[0,1]
\end{equation}
with $\sigma^2$ the statistical variance and $\nabla^{2}_{5}$ the five point operator~\cite{laplacian}. If the blur is detected (i.e., $\varepsilon_{blur} = 0.01$), a loop is launched to check for the disappearance of the fuzziness in the next $B$ frames (i.e., $B=14$).

When faced with long episodes of blur (many consecutive fuzzy frames), a bigger $B$ value is recommended. However, one usually look for low $\varepsilon_{blur}$ values, corresponding to a higher quality standards.
This module record and save all the shift correction parameters as the rejected blurry frames. \newline

\noindent\textbf{Aggregate segmentation and quantification}. After aggregate image normalization and data check, we used a fixed 0.5 threshold to separate the aggregates from the background. Next, we labeled the segmented aggregates to extract features (i.e., count, area and centroid) using the 'label' and 'regionprops' functions from the 'scikit-image' library. To consider that a given labeled aggregate is phagocytosed by a cell, we checked every two consecutive frames if the following conditions are met: the change in the size of the labeled aggregate (decrease by half) and its centroid movement ($0.7\mu m$~$\approx$~$7~pixels$). Finally, all aggregates' features for each time point are reported/saved.\newline

\noindent\textbf{Scene instance-level cell segmentation and tracking modules.}

\noindent\textit{DL and IDL approaches:} U-Nets used four depth levels. In the down-sampling pass, for U-Net and Attention-U-Net, each depth level had a duplication of the following sequence: 2D convolution layers (Conv2D) with 3x3 filters, 2D batch-normalization and leakyReLU, then, 2-factor max-pooling. BiO-Net used a duplication of the following sequence: Conv2D, 2D batch-normalization and ReLU. This sequence is followed by Conv2D, ReLU, 2D batch-normalization and 2-factor max-pooling. The results of each depth level are connected to the symmetrical depth of the decoder as 'skip' connections. 

The midsection (bottleneck) for U-Net and Attention-U-Net was composed of a duplication of the sequence: Conv2D, 2D batch-normalization, leakyReLU. The BiO-Net bottleneck was composed of Conv2D, ReLU, 2D batch-normalization, Conv2D, ReLU, 2D batch-normalization, 2D transposed convolution, ReLU, 2D batch-normalization.

In the up-sampling pass, each depth level used up-sampling with a scale factor of two, and then the skip connection is concatenated differently for each model. In U-Net, it is directly concatenated along the first dimension with two times: Conv2D, 2D batch-normalization and leakyReLU. The Attention-U-Net passed the up-sampled signal through: Conv2D, batch-normalization, and leakyReLU. Then, the attention module (see details~\cite{attunet}) processes the resulting signal and the skip connection. This result is concatenated with the skip connection along the first dimension and passed through the sequence: Conv2D, 2D batch-normalization and leakyReLU. The BiO-Net used the same U-Net decoder module, by only replacing leakyReLU with ReLU. In addition, batch-normalization comes after the ReLU activation function.

For all U-Nets, the output was a single channel image (after Conv2D followed by a sigmoid function). We used the same 2D convolution layers for the encoder and decoder. DL U-Nets contains a (64, 128, 256, 512) sequence of layers for each depth level with a midsection of 1024 layers. IDL U-Nets involved (24, 48, 96, 192) series of layers for each depth level and a midsection of 384 layers. For the BiO-Net, the default 1 iteration and a multiplier of 1.0 are used.

\begin{figure}[H]
    \centering
    \includegraphics[width=0.7\textwidth]{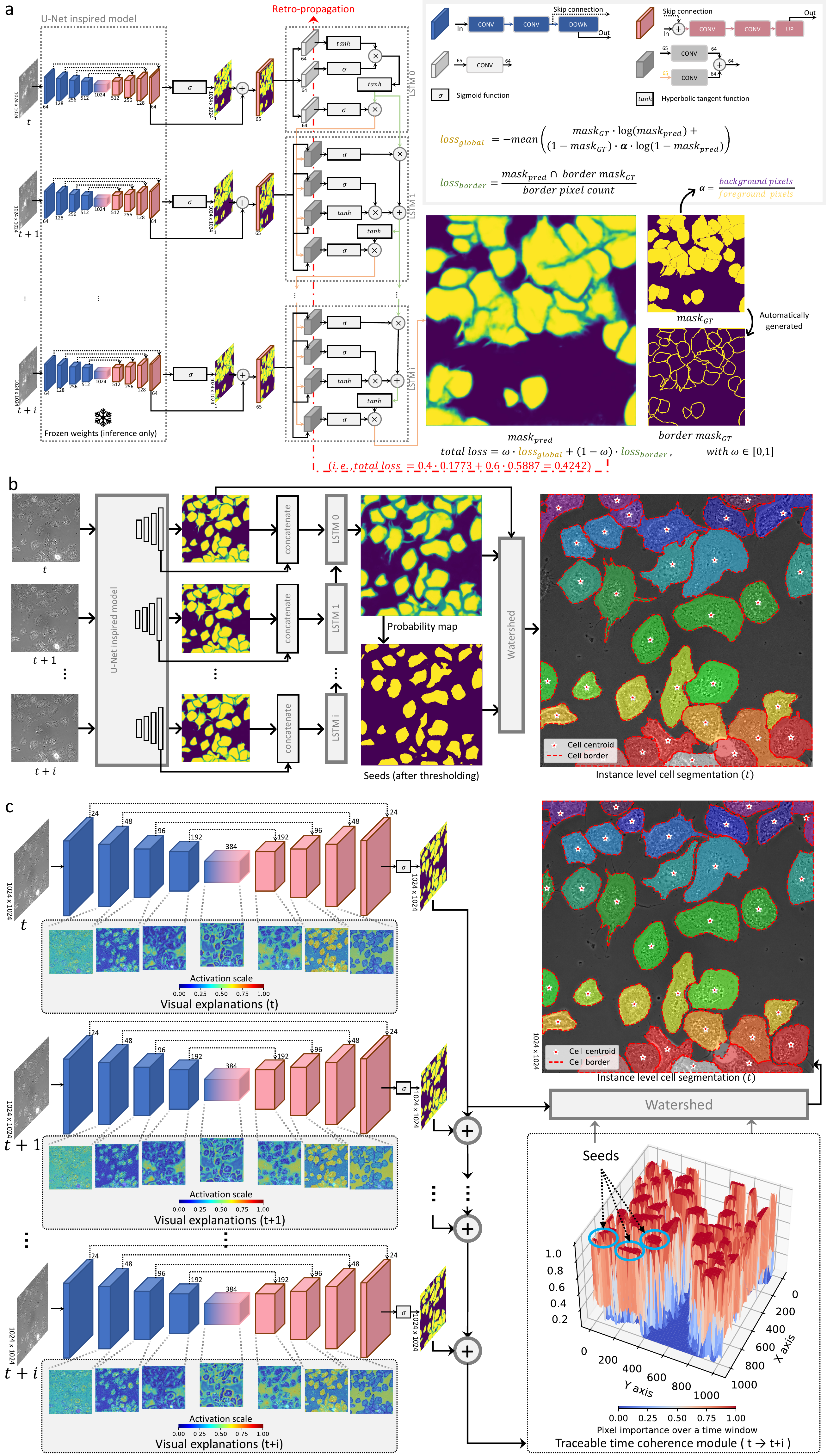}
    \caption{\textbf{Detailed DL and IDL architectures for cell instance segmentation. (a)} Detailed architecture of the segmentation module during the training phase: custom loss functions (global and local) were used during the retro-propagation on the LSTM modules. \textbf{(b)} Detailed inference phase combining U-Net like architectures, LSTM modules and watershed for instance-level cell segmentation. \textbf{(c)} The details of the explainable segmentation module that contains: (i) light U-Net like models attached to a visualization module applied for each time point, (ii) time coherence module (TTCM) that extracts cell seeds, and (iii) watershed module that combines all signals for a full separation.}
    \label{fig4}
\end{figure}

\begin{figure}[H]
    \centering
    \includegraphics[width=\textwidth]{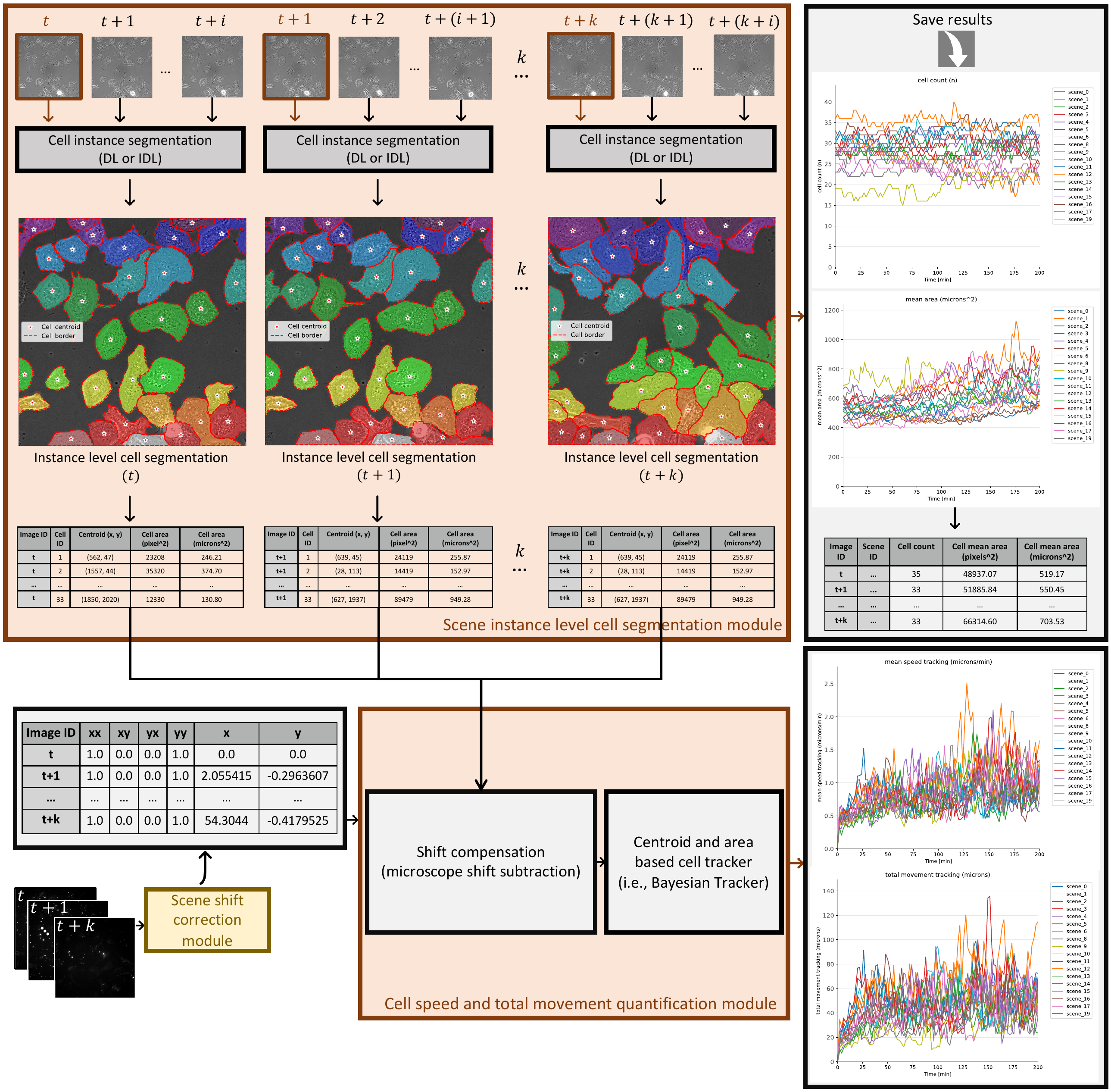}
    \caption{\textbf{Scene cell instance segmentation and tracking.} The scene instance-level segmentation module can use the DL module (Fig.\ref{fig4}.b) or the IDL module (Fig.\ref{fig4}.c) for a scene cell instance segmentation. This module quantifies cell count, area and coordinates for each frame. Cell speed and total movement quantification loads the scene cell features (frame id, centroid, area). Cell centroids are corrected using scene shift correction module (Fig.\ref{fig3}.b). For cell speed and total movement quantification, any tracking algorithm (i.e., Bayesian Tracker) can be applied on the corrected cell features. Results of all complementary modules are saved in an open-source format (i.e., CSV).}
    \label{fig5}
\end{figure}

\begin{figure}[H]
    \centering
    \includegraphics[width=0.9\textwidth]{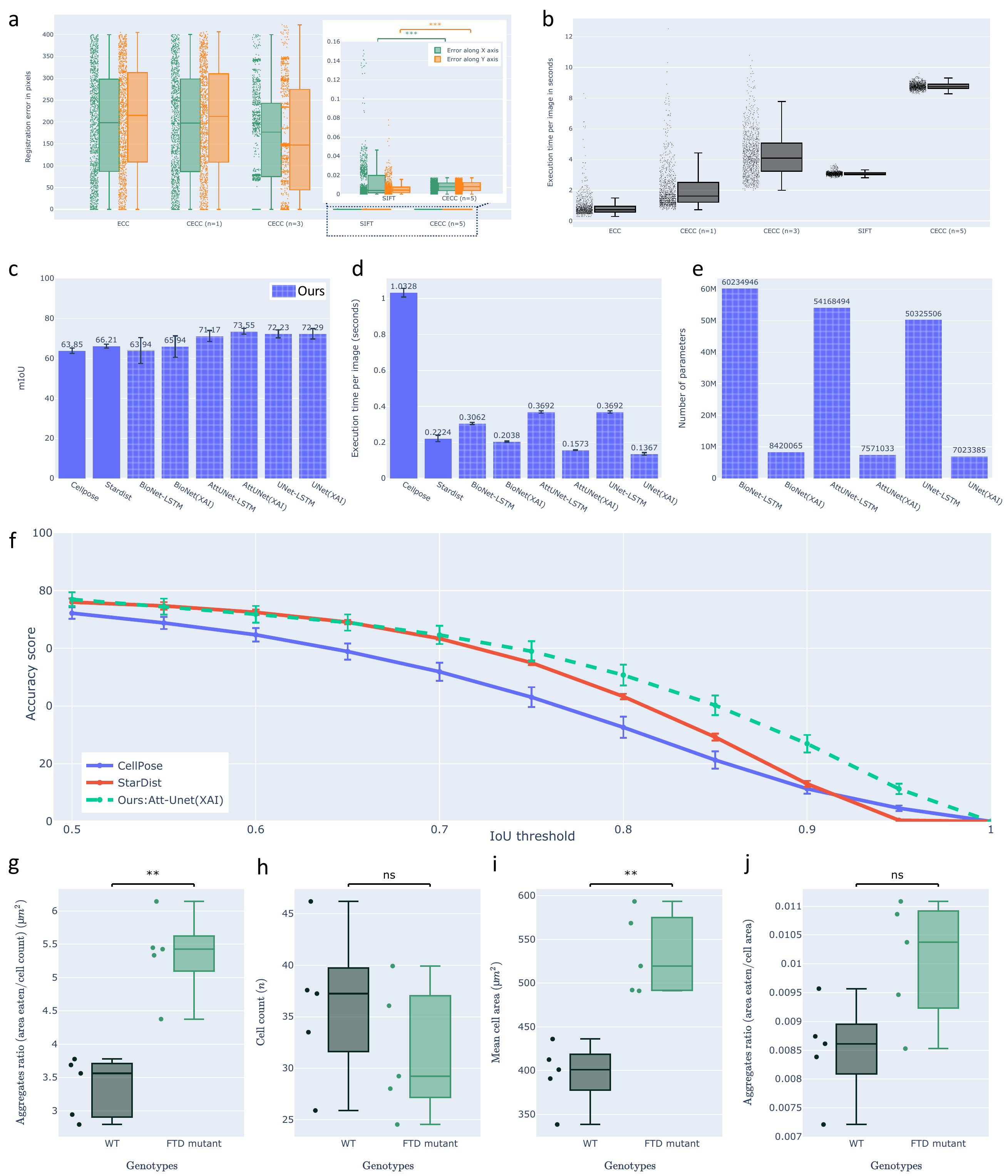}
    \caption{\textbf{Quantitative performance evaluation of the CECC module, DL/IDL cell instance segmentation module and the phagocytic activity of microglial cells in FTD context.} \textbf{(a)} The performance and \textbf{(b)} execution time cost of registration methods ECC, CECC (n=1, 3, 5), and SIFT were evaluated on 1000 randomly shifted frames ($x/y\pm400px$ shift for $2048^2px$ frame). CECC (n=5) achieved the best results with an x/y mean error of $0.008\pm0.004$, outperforming SIFT. Our cell detection approach was evaluated against Cellpose and Stardist on a 165-image test set, using a 5-fold cross-validation/testing approach to compute \textbf{(c)} mean Intersection over Union (mIoU): sum of IoU of the predicted cell masks divided by the ground-truth cell count; \textbf{(d)} the mean execution time cost per image; \textbf{(e)} number of parameters for DL and IDL approaches; \textbf{(f)} the accuracy $(0.5\geq IoU\geq1)$ of our best performing approach 'Att-Unet(XAI)' were computed. Additionally, \textbf{(g)} the amount of TDP-43 aggregates internalized per cell; \textbf{(h)} the number of cells in the assay: cell count; \textbf{(i)} the size of the cells: mean cell area and \textbf{(j)} the amount of TDP-43 internalized per cell surface unit. Statistical tests were conducted using the Mann-Whitney-Wilcoxon test with ns (p-value $\geq$ 0.05), ** (p-value under 0.01), and *** (p-value under 0.001).}\label{fig6}
\end{figure}

\begin{figure*}[ht]
    \centering
    \includegraphics[width=\textwidth]{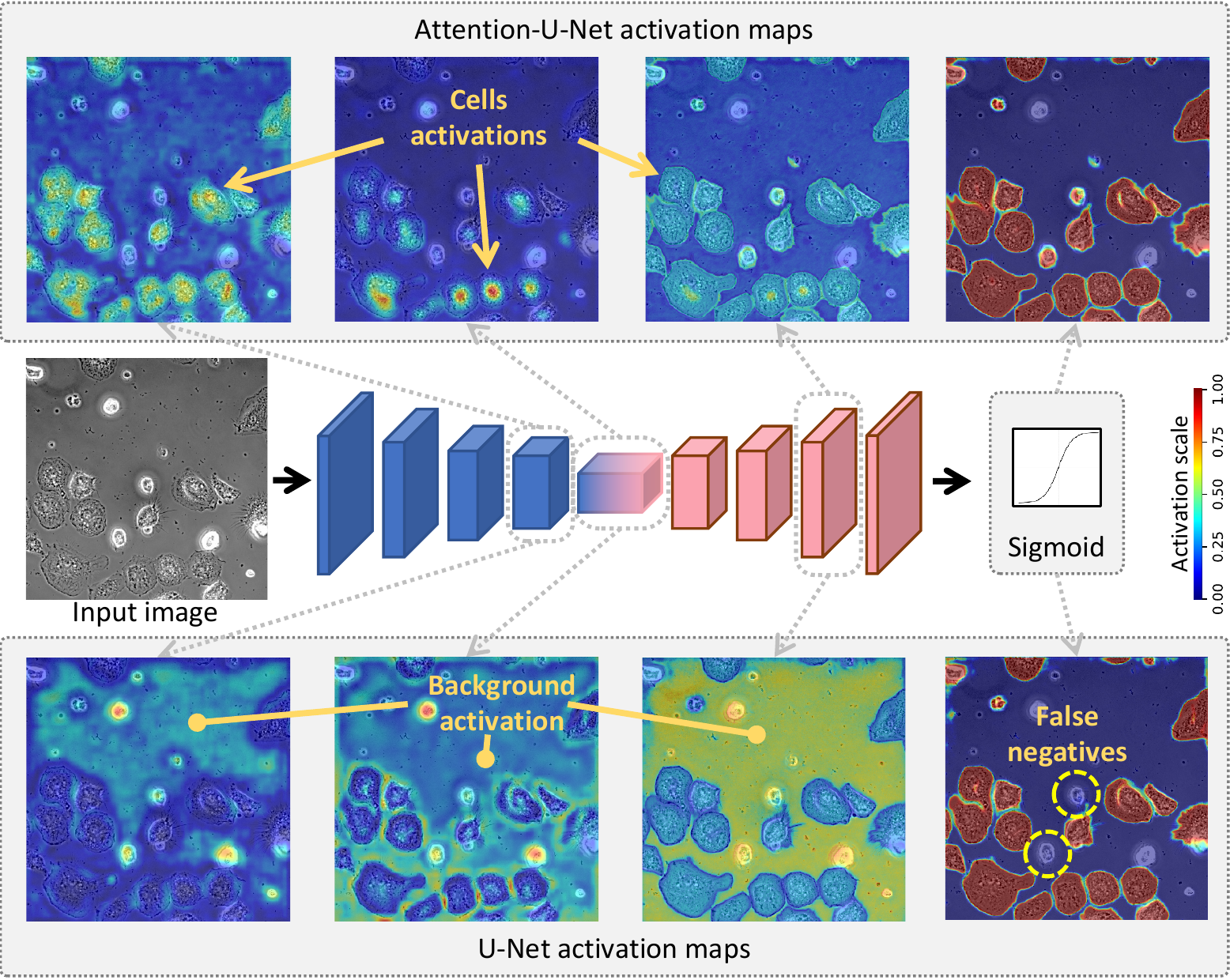}
    \caption{\textbf{Visualization of the features learned by U-Net versus Attention-U-Net.} U-Net model focused more on the background features since the background texture is easier to model compared to the texture of the cells. Consequently, it led U-Net to produce false negatives. On the other hand, Attention-U-Net uses the attention mechanism, which makes it focus on the cells' texture, thus, yielding fewer false negatives compared to U-Net.}\label{sup_fig_1}
\end{figure*}

\begin{figure}[H]
    \centering
    \includegraphics[width=0.8\textwidth]{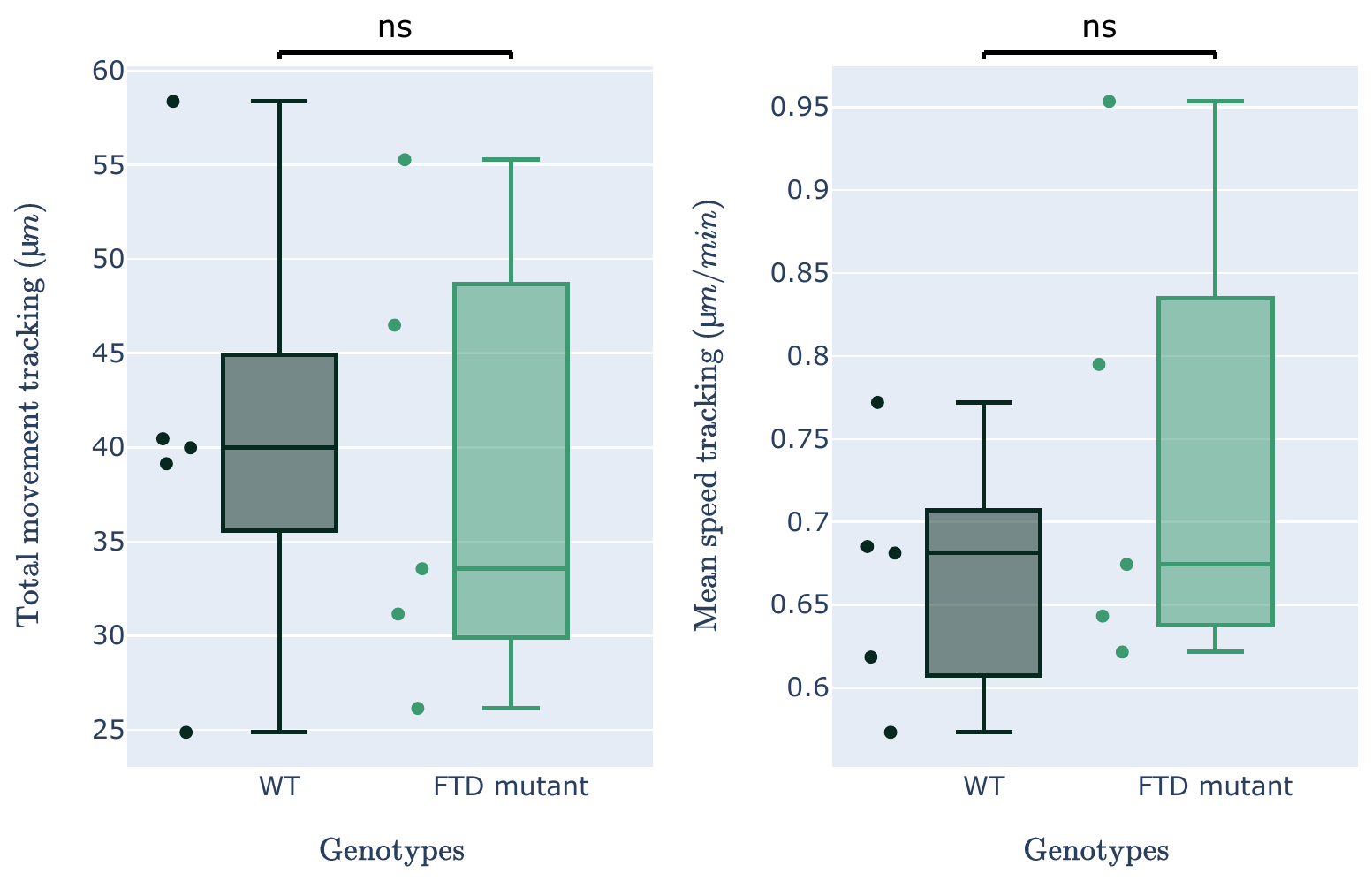}
    \caption{\textbf{Additional quantitative results of FTD-mutant versus WT microglial cells.} \textbf{(right)} the quantification of total cells movement. \textbf{(left)} the quantification of the cells' mean speed. (statistical test: Mann–Whitney–Wilcoxon, ns: p-value $\geq$ 0.05)}\label{sup_fig_2}
\end{figure}

\begin{figure}[H]
    \centering
    \includegraphics[width=\textwidth]{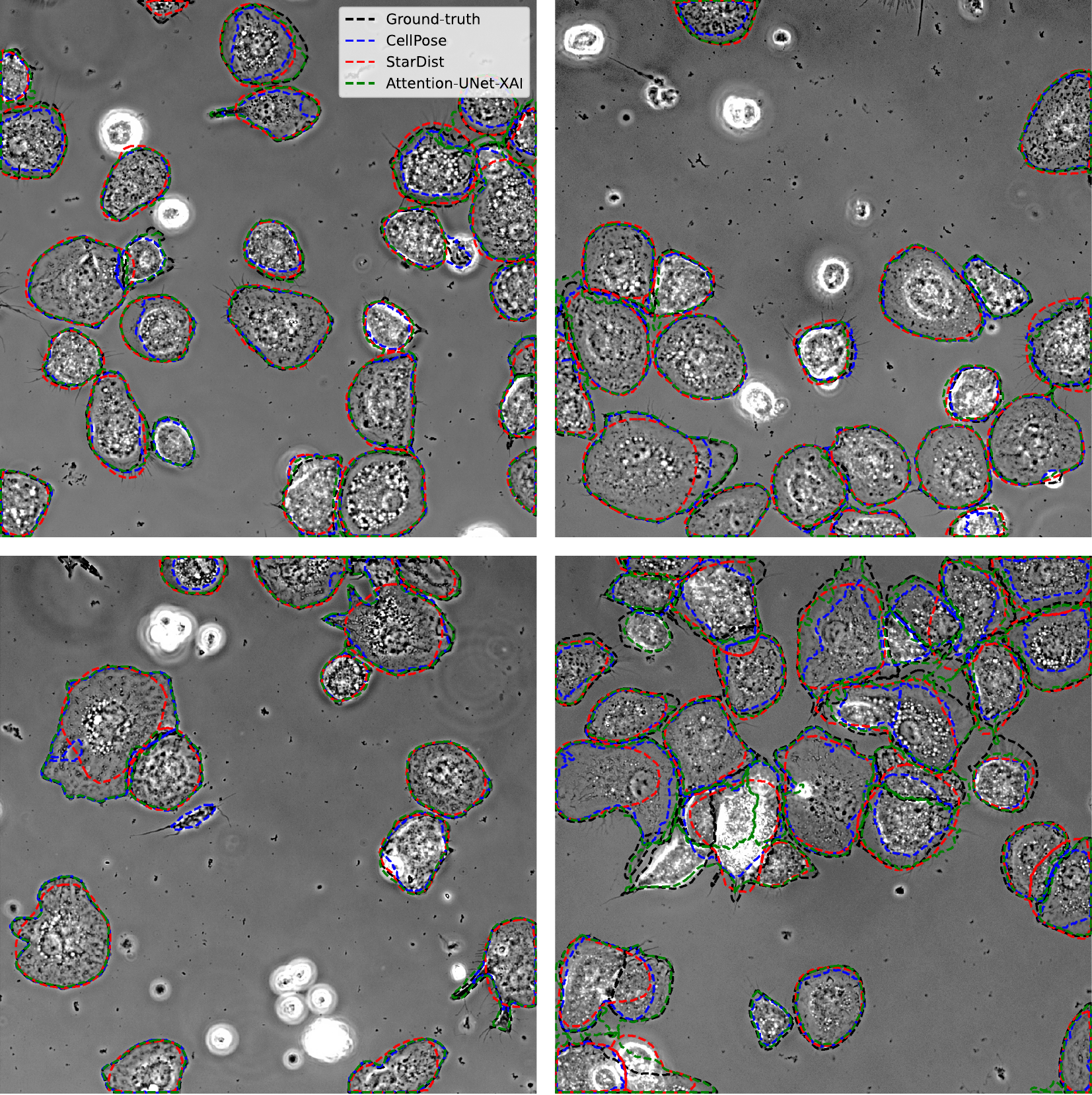}
    \caption{\textbf{Instance-level cell segmentation evaluation:} Our qualitative analysis demonstrates that the Attention-UNet(XAI) model performs favorably in comparison to Cellpose and Stardist, particularly in handling challenging cell shapes. This evidence suggests that our approach adapts effectively to the complexities associated with cell morphology, showcasing its potential to provide a competitive alternative to existing state-of-the-art methods. Notably, in extreme cases where cells are agglomerated and still in suspension (as seen in the white cluster of cells at the bottom right), all methods, including our own, struggle to achieve accurate segmentation. Such instances highlight the need for further refinement and optimization to address these challenging scenarios effectively.}\label{sup_fig_3}
\end{figure}

\begin{figure}[H]
    \centering
    \includegraphics[width=\textwidth]{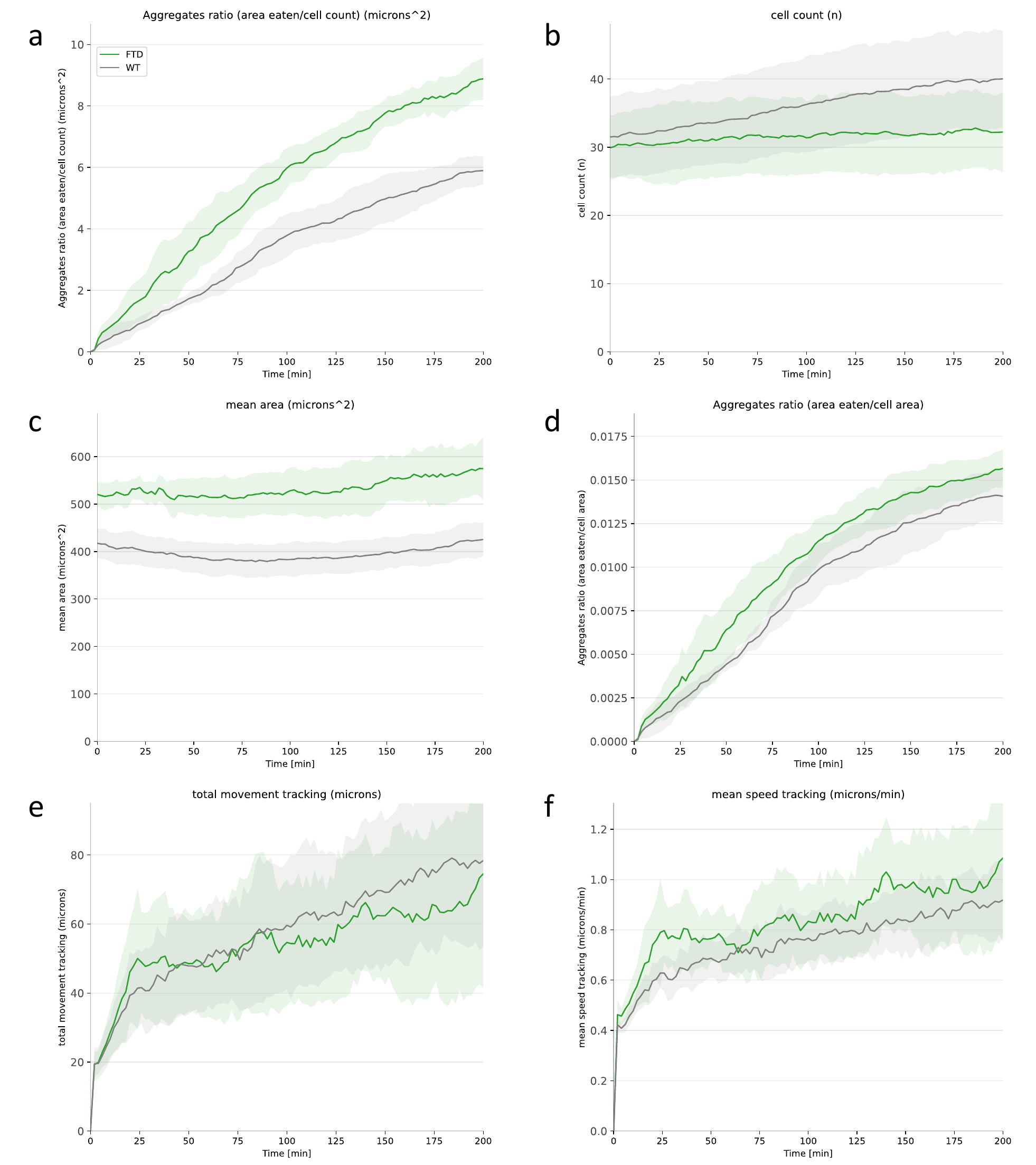}
    \caption{\textbf{Comparative analysis of phagocytosis metrics over time for WT and FTD Groups.} This figure presents a time-course comparison of phagocytosis-related parameters between WT and FTD groups. The panels display (a) aggregate area consumed by cells, (b) cell count, (c) mean cell area, (d) cell surface area consumption, (e) total cell movement, and (f) cell speed over time. This analysis highlights the dynamic differences in phagocytosis between the two conditions and enables a comprehensive understanding of their distinct behaviors.}\label{sup_fig_4}
\end{figure}

\begin{figure}[H]
    \centering
    \includegraphics[width=\textwidth]{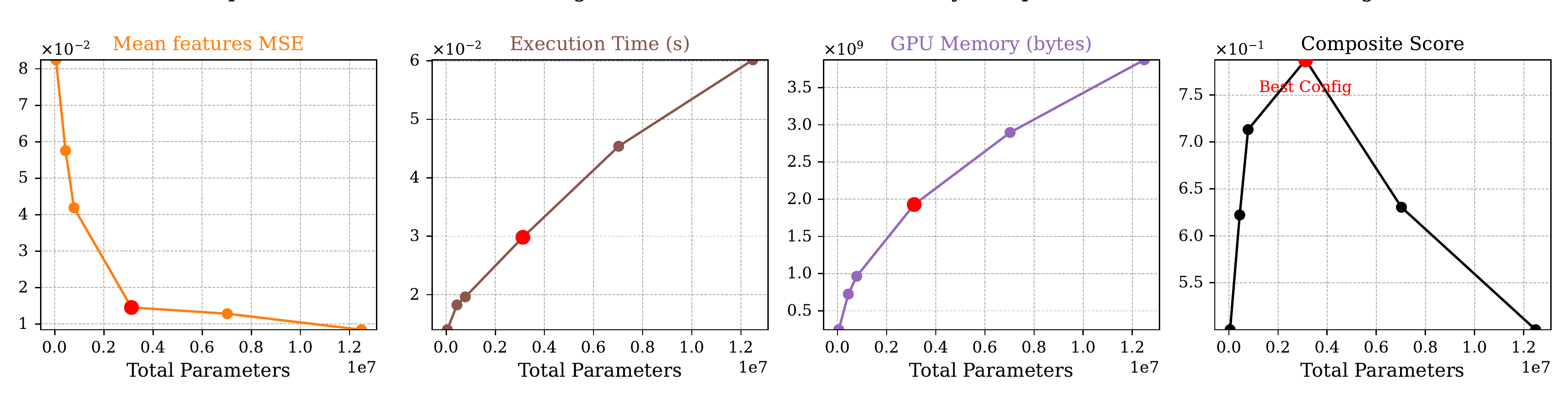}
    \caption{\textbf{Automated DL model optimization using feature maps: balancing feature map quality (MSE) and execution time (sec) in Unet Models} We illustrate the evaluation of various quantitative metrics including (from the left) the Mean Squared Error (MSE) of feature Map signal quality compared to 30M Unet model (reference model), execution time (seconds), and GPU memory utilization (bytes). A composite score is calculated using the formula: \( \alpha \times \text{time} + \beta \times \text{memory} + \gamma \times \text{MSE} \), all metrics are min-max normalized (to make them range between 0 and 1). If a lower value indicates better performance, the normalized metric was adjusted to 1 minus its value. With \( \alpha = 0.5 \), \( \beta = 0 \), and \( \gamma = 0.5 \). The red point indicates the best trade-off achieved between execution time and feature map signal quality, presenting the ideal parameters for the Unet model.}\label{sup_fig_features}
\end{figure}

\begin{figure}[H]
    \centering
    \includegraphics[width=\textwidth]{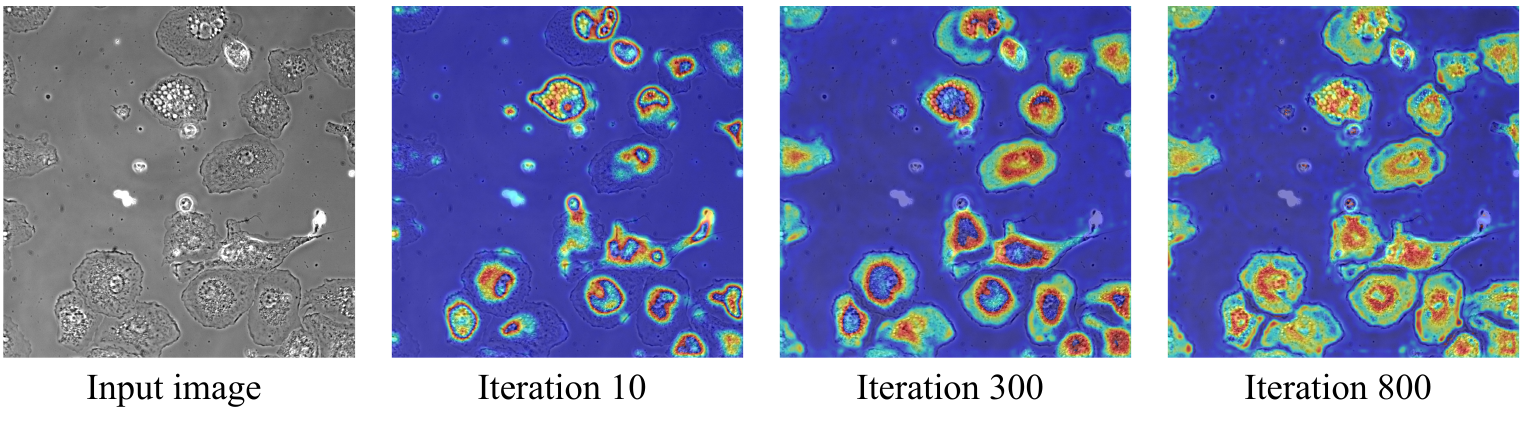}
    \caption{\textbf{Progressive learning visualization in AttUNet deep learning model training.} We qualitatively illustrates the key stages in the learning process of our deep learning UNet model, as revealed through heatmaps (mean feature map), which are pivotal in demonstrating the model's evolving focus. At the 10 iteration mark, the model starts to identify cell textures, separating cells from the background. Progressing to 300 iterations, it hones in on intracellular components, delineating cell boundaries and nuclei. By 800 iterations, the model exhibits advanced recognition capabilities, identifying cells with partial visibility and precisely differentiating between individual cells. These visualizations are critical in building trust with neuroscientists by providing transparent insights (refer to section~\ref{sec:interpretability-heatmap}) into the model's learning dynamics.}\label{DL_evo}
\end{figure}

\begin{figure}[H]
    \centering
    \includegraphics[width=0.48\textwidth]{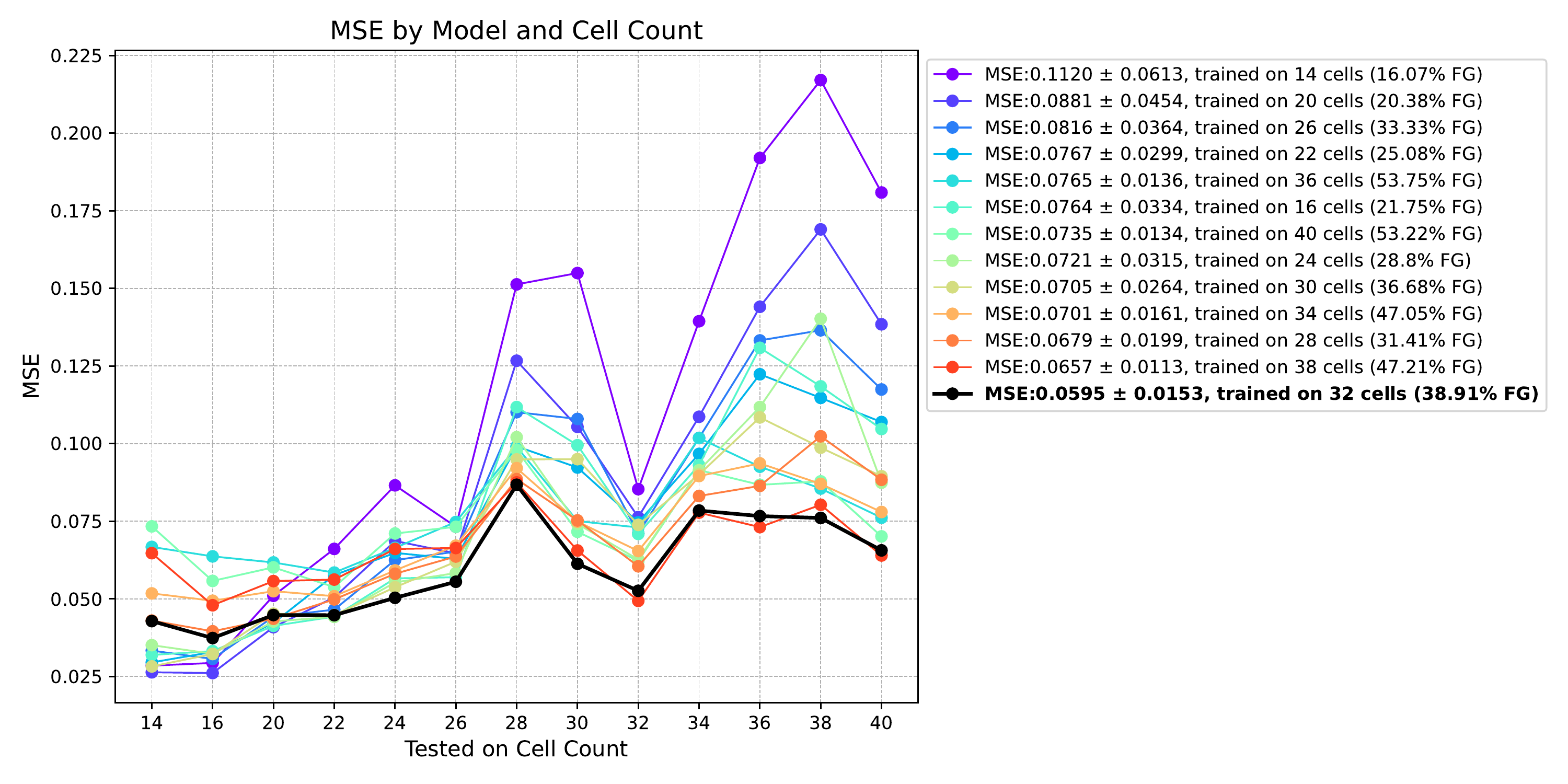}
    \includegraphics[width=0.48\textwidth]{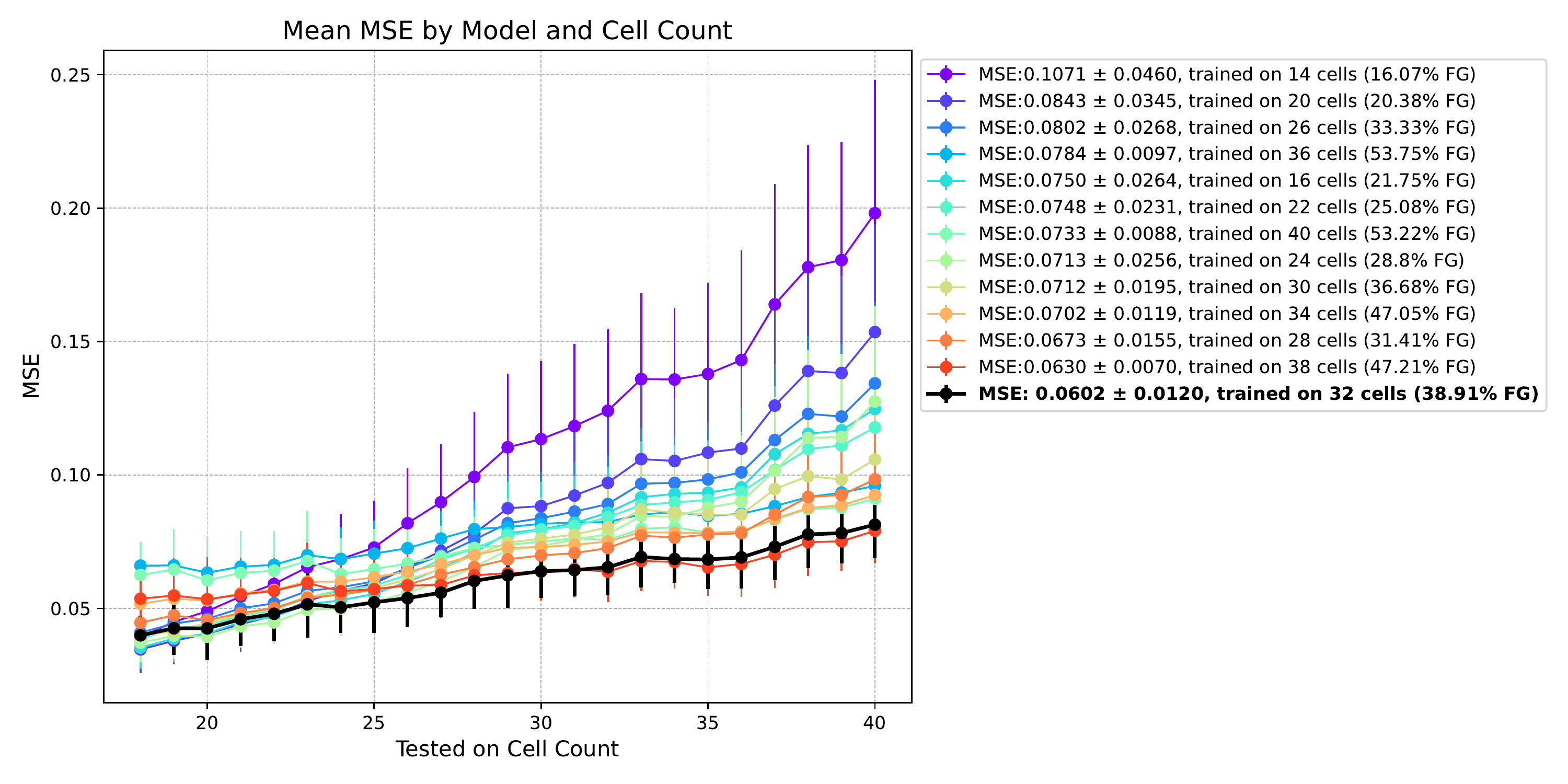}
    \caption{\textbf{AttUnet(XAI) sensitivity assessment on test images cell quantities per condition.} This figure juxtaposes the results of two distinct test setups for evaluating model performance. On the left, we present results from our sensitively analysis framework, where only three image per cell count was used for training, validation and testing, significantly reducing the data requirement. On the right, we illustrate the outcomes using 100 test images per condition across 23 conditions on the trained models, involving a 2300 images with varying cell counts. The MSE (lower is better) calculated between the model-generated probability maps and the corresponding ground truth binary masks. The top-5 performing models suggests practical guidelines, such as prioritizing the annotation of images with cell counts between 28 and 38 and a foreground ratio of 31\% to 47\%. These results highlight the efficacy of our sensitivity assessment framework in identifying crucial image characteristics that impact performance, guiding annotators towards more strategic and efficient annotation processes with minimal data, with the ability to investigate the model's behavior in controlled settings, without major time or computational burdens. Trainig the 13 different cell counts models required less than 30 minutes in total, employing a single 8GB GPU (NVIDIA RTX 2080).
}\label{sensitivity_dl}
\end{figure}

LSTM modules (described in Fig.\ref{fig4}.a) were connected to the frozen U-Nets (forward-pass only). The highest encoder depth convolution results (64x1024x1024) were concatenated with the prediction image (1x1024x1024) and passed to the $LSTM_0$ when the given frame is the first one in the chosen time-window (successions of frames), otherwise to $LSTM_i$.

The TTCM (presented in Fig.\ref{fig4}.c) concatenated the probability maps from U-Nets. These results were then normalized by the number of the time points. Seeds were finally extracted using a high thresholding (i.e., 0.9), corresponding to selecting the pixels presented in most of the frames of the time-window.

The visual explanation module is connected to the XAI U-Nets. Each depth level (encoder and decoder) output was extracted before the mean activation heat map was computed along axis 1. The resulting image was scaled to match the input image dimensions (1024x1024) using the 'resize' function from the 'PIL' Python library.\newline

\textit{DL training phase:} Building on the established conventions for U-Nets in semantic segmentation, our model introduces a critical enhancement with the alpha factor (\(\alpha_i\)) for calculating the global loss. This factor is dynamically computed for each training image, allowing our binary cross-entropy loss function to adapt to the unique ratio of background to foreground pixels in each image's ground truth data. For a given image \(i\), if the cellular density is low, the alpha factor increases the weight of the cell pixel class within the loss function. Thus, the global loss for image \(i\) is defined by incorporating the alpha factor, \(\alpha_i\), to ensure that the loss is representative of the actual class imbalance on a per-image basis. This approach shares conceptual similarities with methodologies such as NeuRegenerate's density multiplier\cite{NeuRegenerate}, which adapts model behavior to address the tile-stitching artifacts. In NeuRegenerate's case, this adaptation is based on the overlap between synthetic and real inputs in a 3D volumetric context, particularly when computing the reconstruction loss within a generative adversarial network setting. Our alpha factor, however, is specifically tailored for 2D image segmentation, enhancing sensitivity to the nuances of each training image, presenting a substantial improvement in how class imbalances are addressed in the model. Where the global loss is formulated as:
\begin{equation}
    loss_{global} = -\frac{1}{I}\sum_{i=1}^{I} \left[gt_{i}\cdot log(pred_{i}) + (1-gt_{i})\cdot\alpha_{i}\cdot log(1-pred_{i})\right]\label{eq:bce_mod}
\end{equation}
With $\alpha_i$ is the ratio between the number of background and foreground pixels from the ground~truth for the \(i\)-th image, $I$ represents the number of training images, $gt_{i}$ the ground truth binary mask, $pred_{i}$ the model prediction. In order to optimize the speed, the 'PyTorch' library is used to flatten the masks. Then the loss function is computed directly on the GPU, and reducing the delay between the forward and the backward passes.

In a first phase of the DL training, all U-Nets were trained using 5-fold cross-validation and testing while using: (i) $loss_{global}$ for retro-propagation (see equation~\ref{eq:bce_mod}); (ii) Adam optimizer; (iii) $10^{-4}$ learning rate and (iv) batch size of one. After twenty epochs, the best model was saved for each validation fold based on its $loss_{global}$ score on the validation set, then tested on the test set.
In order to take into account the cell border and to reduce the training time, border masks were automatically generated for the dataset in the following manner: cell binary mask was dilated using the 'binary\_dilation' function from the 'scipy' library for two iterations, the pixels of the original mask was subtracted (keeping only the borders after dilation), then the border mask was dilated for 4 successive iterations (see Fig.\ref{fig4}.a). We defined the border loss as:
\begin{equation}
    loss_{border} = \frac{1}{I}\sum_{i=1}^{I}\frac{|pred_{i} \cap gt\_border_{i}|}{|gt\_border_{i}|}
\label{eq:border}
\end{equation}

Let \( I \) represent the number of training images. For each \( i^{th} \) image, \( gt\_border_{i} \) is the set of pixels constituting the automatically generated ground-truth border, while \( pred_{i} \) is the set of pixels where the model predicts a border. If the model's prediction, \( pred_{i} \), does not intersect with any of the true border pixels from \( gt\_border_{i} \), the intersection is empty and thus \( loss_{border} = 0 \). Conversely, if every pixel in \( gt\_border_{i} \) is also in \( pred_{i} \), indicating a total overlap, then \( loss_{border} = 1 \).

In a second phase of the DL training, the parameters of the U-Nets were frozen, inhibiting any back-propagation. Subsequently, the U-Nets were linked to LSTM modules that functioned with a two-time point window at a time ($LSTM_0$ and $LSTM_1$), permitting back-propagation to modify only the LSTM parameters (refer to Fig.\ref{fig4}.a). In our training, the loss function was a combination of equations \ref{eq:bce_mod} and \ref{eq:border}:
\begin{equation}
    total\_ loss = \omega\cdot loss_{global} + (1-\omega)\cdot loss_{border},~~~\omega\in[0,0.5] \label{eq:total}
\end{equation}
The coefficient \( \omega \) is pivotal for controlling the weightage given to the global versus the border loss. From preliminary experimentation, \( \omega = 0.4 \) was discerned to be a balanced choice, thereby augmenting cell separation. For instance, a lower \( \omega \) value of 0.1 improved precision but slightly detracted from recall. It is noteworthy that values of \( \omega \) exceeding 0.5 jeopardized cell separation, eliciting declines in both precision and recall metrics. As a consequence, the scope of \( \omega \) was confined to the interval [0, 0.5]. This value of \( \omega \) not only emphasizes cell borders—a crucial factor for our cell detection quality—but also ensures the retention of important global features of the image, such as demarcating the foreground from the background.

The LSTM modules were trained using the 5-fold cross-validated U-Net frozen models, $total\_loss$ with $\omega=0.4$ (see equation~\ref{eq:total}), Adam optimizer, $10^{-4}$ learning rate and a batch size of one. After twenty epochs, the best model was saved for each validation fold based on its $total\_loss$ score on the validation set and then tested on the test set.\newline

\noindent\textit{IDL training phase:} U-Nets were trained using 5-fold cross-validation and testing, $loss_{global}$ for retro-propagation (see equation~\ref{eq:bce_mod}), Adam optimizer, $10^{-4}$ learning rate, batch size of one. After twenty epochs, the best model was saved for each validation fold based on its $loss_{global}$ score on the validation set, then tested on the test set.\newline

\noindent\textit{DL/IDL inference phase:} For DL we used UNets+LSTM and for IDL we used UNets+TTCM. These modules combination produced time-series-based probability maps (high-values: cells, low-values: background and cell borders). Then, cell seeds (centroid coordinates) were extracted after 0.9 thresholding. Watershed method combined the probability map as a distance map, cell centroids as seeds and the binary mask (U-Nets predictions after 0.5 thresholding) as a foreground delimiter (see Fig.\ref{fig4}.b, Fig.\ref{fig4}.c). Moreover, the execution time evaluation (during inference) presented in Fig.\ref{fig6}.d was performed using the following hardware 8 cores i7 9700K CPU, 16GB RAM, Nvidia RTX2080 GPU and Samsung 970 PRO SSD.\newline

\noindent\textit{Data input size to all models:} Each frame was resized to 1024 x 1024 pixels before being input into the model. This decision was taken to strike a careful balance between maintaining high resolution for effective model performance, storage scalability and ensuring computational efficiency. The original resolution of 2048 x 2048 pixels was reduced to fit most hardware capabilities while still preserving sufficient detail for the model’s tasks. Indeed, we did not utilize a tiling strategy for the frames; each was processed in its entirety at the reduced size (1024 x 1024 pixels). This approach eliminates concerns about tiling overlap and its potential implications.\newline

\noindent\textit{Point2Cell annotation tool:} Point2Cell integrates a pair of UNet models for distinct purposes. The first UNet model predicts binary cell masks, while the second focuses on cell density estimation. Cell density maps are generated using the Distance-map library\cite{distance_map}, which employs a geodesic distance measure from a Python library to create 2D distributions around cell centroid coordinates, or seeds. In this system, the Distance-map library uses the Euclidean distance metric, adjusted by a linear alpha parameter, and applies it to cell centroids. These distributions are refined using binary masks from the first UNet model, yielding accurate cell density maps. Both UNet models are trained from scratch, with a batch size of one. Optimization of parameters is done using the RMSprop optimizer, at a steady learning rate of $10^-4$. The training employs early stopping at 10 epochs and is capped at 200 epochs. Point2Cell also incorporates a user-interactive cell seeding feature for image annotation, where users manually identify each cell in an image. This feature includes options like undo, reset, and save, enhancing annotation efficiency. This manual seeding is essential for accurate annotations. After manual seeding, Point2Cell uses its trained models to generate pseudo-cell masks and density maps. These, along with the user-provided cell seeds, are processed through a watershed algorithm~\cite{watershed1, watershed3, deepwatershed}. This method effectively isolates and labels individual cells. In tests, Point2Cell showed greater annotation accuracy than polygon-based annotation. Using 10 images from the HeLa cells dataset from CTC, it achieved a Dice score of 94.97\% in just 14.6 seconds. In comparison, the polygon-based 'labelme' tool scored 91.3\% but took much longer, at 96.1 seconds. Point2Cell's efficiency is highlighted by its speed, being about six times faster than 'labelme', and its superior precision, with a 3.64\% higher Dice score. Point2Cell's source code is publicly available on GitHub. It streamlines cell annotation, requiring only single-click input from the user for highly precise, pixel-level cell annotations.\newline

\noindent\textit{Cell tracking:} The Bayesian Tracker (btrack)\cite{btrack1,btrack2} Python library was used to track cells over time. It used the centroid and area to form cell tracks. Only the tracks with at least 100 min long were kept. Speed was computed at each time point (mean cell displacement divided by time unit), quantifying speed over time, and then, mean speed over a whole sequence was computed and reported. A similar approach was used to compute total displacement over time and for the whole sequence.\newline

\noindent\textbf{AttUNet(XAI) and UNet(XAI) as pre-trained models:} The training of our models on the full dataset for 20 epochs took approximately 30 hours for the Att-Unet(XAI) and 20 hours for the Unet(XAI) using the Nvidia GPU: Tesla V100-SXM2-32GB. We consider this to be quite efficient given the high input resolution (1024 x 1024 pixels) and the complexity of the task the model is designed to perform. We trained the two models from scratch using the entire dataset so that the community will have the possibility to use it as is on similar data or fine-tune it for similar tasks. We used 3 FTD + 3 WT experiments (around 120 sequences, 22,412 images, and 22,412 masks) for training, 1 FTD + 1 WT experiment (around 40 sequences, 7,323 images, and 7,323 masks) for validation, and 1 FTD + 1 WT experiment (around 40 sequences, 6,761 images, and 6,761 masks).\newline

\noindent\textbf{Automated DL model optimization using feature maps:} We used Unet default configuration as our primary reference. With this model in place, we used objective metrics such as MSE. This involved considering the average feature maps generated from a non annotated set, where each image was processed both by the reference model and its smaller versions.

As a concrete use case, we maintained uniform parameters: the same random seed, identical training/validation splits, and a consistent number of training epochs. As our benchmark, the U-Net model came equipped with 30M trainable parameters (default configuration). Simultaneously, we ventured into training its more compact counterparts. While these variants had fewer convolution layers, their foundational architecture mirrored the original (same number of max-pooling and up-sampling stages), with parameter counts ranging from 12.5M to a mere 49K.

Transitioning to the evaluation phase using the test set, we compared the output feature maps of the reference 30M model against those of its compact versions. Using metrics like MSE, we derived an average comparative score for each model. Concurrently, we monitored two computational metrics: inference times and GPU memory consumption.

For visualization, we plotted all these metrics, making them easily digestible. Our comprehensive assessment also included a composite score, formulated as:
\begin{equation}
    composite\_score = \alpha \cdot execution\_time +  \beta \cdot memory +  \gamma \cdot feature\_map\_signal\_quality
\end{equation}
Where  \( \alpha + \beta + \gamma = 1 \), and each one ranges between [0, 1]. Execution time performance (execution time per image in seconds), memory performance (memory footprint of the model + 1 inferred image), and feature map signal quality MSE are all min-max normalized (to make them range between 0 and 1). If a lower value indicates better performance, the normalized metric was adjusted to 1 minus its value.

As an example, we showcase plots (refer to Fig.\ref{sup_fig_features}) representing the non-normalized MSE, execution time, and GPU memory. Along with the composite score, where \( \alpha =0.5, \beta =0, \gamma = 0.5 \). We present the best trade-off between execution time and feature map signal quality using the red point.

In summary, this approach not only pinpoints a model's ideal size but also establishes a harmonious balance between feature map signal quality and execution speed. This automation not only elucidates our methodology but also empowers users to adjust model parameters according to their unique needs, balancing model complexity, feature map signal quality, memory footprint, and processing time.\newline

\noindent\textbf{Coherence score boundary cases:} In our methodology, the coherence score gauges the continuity of a cell’s visibility and tracking across successive frames. When a cell first appears in our field of view at frame \( n \), we look at the following frames within a specific time window (frame \( n + \text{time window} \)) to evaluate and quantify its movements. This contributes to the determination of the cell instance mask for frame \( n \). However, there are exceptional boundary cases where the coherence score might not truly represent cell activity:
\begin{itemize}
    \item If a cell becomes visible within the field of view after the final frame minus the time window (\( \text{last frame} - \text{time window} \)), it may be assigned an inappropriately low coherence score. It's important to note that we've taken measures to prevent such edge cases by recording an additional 30 minutes, totaling 7 hours and 30 minutes. Since we only require 7 hours, this ensures we remain within a safe margin.
    \item Cells that move in and out of the field of view frequently, in what we term 'field of view border kill', can also skew the coherence score.
\end{itemize}

Moreover, it should be noted that dead cells may exhibit high coherence scores because they persist in the same location, regardless of their lack of movement. Currently, our approach does not reduce the coherence score to reflect this immobility. In our research, we find it unnecessary to account for this—as neurologists didn't qualitatively notice dead cells—but it is straightforward to modify the approach to include parameters that would penalize the coherence score for non-movement. While not critical for our study, such an adjustment could improve the precision of analyses in other research areas where distinguishing between live and non-viable cells is crucial.\newline

\noindent\textbf{Microglia primary culture}.
Microglia primary cultures were performed using newborn brains of controls (C57BL6/J), of FTD-mutant animals (line C9orf72-/- or GrnR493X/R493X). Newborn mice brains (less than two days old) are collected by dissection of the skull. Brains are recovered in a 50mL Falcon and mechanically dissociated by gentle pipetting into 5mL of Hank's Balanced Salt Solution (HBSS Thermo Fisher Scientific 14025050). After dissociation, the resulting cell suspension is then centrifuged at 1200rpm for 10 minutes at 4°C. The pellet is re-suspended with culture medium containing DMEM (Thermo Fisher Scientific 31885023), supplemented with 10\% de-complemented calf fetal serum free of endotoxins (HI FBS Thermo Fisher Scientific 10082147), 1\% Penicillin + Streptomycin (Thermo Fisher Scientific 15070063). The cell suspension is cultured in flasks (75 mm2) previously coated with Poly-L-Lysine (SIGMA P4832) for 30 minutes at 37°C (5\% CO2) then washed three times with 1X Phosphate Buffered Saline. The culture flasks are incubated at 37°C (5\% CO2). Fifteen days later, microglia are ready for harvest. Microglia are obtained by light shaking and recovery of the culture medium in a 50mL Falcon. After centrifugation, cells are re-suspended in fresh culture medium and plated.\newline

\noindent\textbf{Phagocytosis assay}. Aggregates of recombinant human full length TAR DNA-binding protein 43 (TDP-43, Abcam ab156345) were conjugated to Alexa Fluor 555  NHS Ester (ester succinimidyl, Thermo Fisher Scientific A20009) at equimolar concentration and deposited on a 35 mm glass-bottom dish (Ibidi, 81218-200) for 2 hours at $37^\circ$C, 5\% CO2. The dish was then washed 3 times with 1X phosphate buffered saline (PBS) and $12.5$ x $10^{5}$ freshly harvested primary mouse microglia (WT, \textit{Grn} KO or \textit{C9orf72} KO) were seeded on top of the fluorescent aggregates in DMEM (Thermo Fisher Scientific 31885023), supplemented with 1\% N2 supplement (Thermo Fisher Scientific 17502048) and 1\% Penicillin + Streptomycin (Thermo Fisher Scientific 15070063). Within 30 minutes after seeding the culture dish was placed in a Zeiss Axio Observer 7 video-microscope at $37^\circ$C, 5\% CO2 and video were acquired at 63X for 7h ($2048\times2048$ images with 0.103$\mu m$ x 0.103$\mu m$ per pixel). For the sake of simplicity, we summarize the steps of data preparation as follows: (i) fluorescent aggregates were deposited onto a glass bottom culture dish and incubated for two hours; (ii) the dishes were washed three times after incubation; (iii) freshly harvested primary mouse microglia wild type (WT) or FTD-mutants were implanted on top of the fluorescent aggregates; and (iv) the culture dishes were placed in a video microscope 30 minutes following seeding, and a video was recorded accordingly.\newline

\noindent\textbf{FTD-mutants versus WT cells}. The results presented in Fig.\ref{fig6}.g, Fig.\ref{fig6}.h, Fig.\ref{fig6}.i, Fig.\ref{fig6}.j and Fig.\ref{sup_fig_2} (extended data) were computed in the following manner: (i) we computed the mean curves of all scenes, where we had 20 scenes maximum per acquisition, and each acquisition is (n=1) and (ii) we computed the mean values from the curves between 0 and 200 min.\newline

\noindent\textbf{Data collection} Imaging of cells/aggregates in 2D+time was performed on a Zeiss Axio Observer 7 video-microscope at 37°C, 5\% CO2 and videos were acquired at 63X for 7h ($2048\times2048$ images with $0.103\mu m$ x $ 0.103\mu m$ per pixel). We used the ZEN Microscope Software v2.6.76.0. We conducted a total of 10 experiments, with 5 using wild-type (WT) cells and 5 using FTD mutant cells. In each of these experiments, we included the cells from 6 pups. Thus, we analyzed cells from 30 pups for WT and another 30 pups for FTD mutant cells, totaling 60 pups across all experiments. Regarding the details provided in the Methods section, the term "20 scenes per acquisition" corresponds to the capture of 20 distinct imaging sequences in each experiment. Each sequence represented 7 hours of continuous imaging using phase-contrast video microscopy, resulting in a collection of 200 unique sequences across all experiments (calculated as 10 experiments multiplied by 20 sequences per experiment). It is not a multiplicative factor of the number of pups but rather the number of sequences per experiment.\newline

\noindent\textbf{Laboratory animals} Mus musculus, C57BL6J, newborn mice were euthanized by decapitation as recommended for rodents up to 10 days of age. They were sacrificed to generate the microglial primary culture, parents were 4 to 8 months old. Mice were kept on a 12h light/dark cycle with food and water available ad libitum. Temperature between 19 and 24°C and humidity between 45\% and 65\%. To do microglial primary cultures, postnatal day one mice pups of both sexes are used and cells from all animals dissected on the same day are only pooled by genotype. As the same occurs for all genotypes it does not impair our differential analysis.\newline

\noindent\textbf{Compliance with essential ARRIVE guidelines} \\
\noindent\textbf{Study design:}
\begin{enumerate}[label=\alph*)]
    \item Control group: Wild type (WT) (C57BL/6JRj), FTD-mutant (line C9orf72-/- or GrnR493X/R493X)
    \item Experimental unit: Litter (each experiment was performed with cells extracted from one litter of pups per genotype)
\end{enumerate}

\noindent\textbf{Sample size:}
\begin{enumerate}[label=\alph*)]
    \item Six pups per experiment, resulting in a total of 60 pups for all experiments conducted in this study.
    \item Sample sizes of n=5 for WT and n=5 for FTD mutants are typical for in vivo studies.
\end{enumerate}

\noindent\textbf{Inclusion and exclusion criteria:}
\begin{enumerate}[label=\alph*)]
    \item Experimental units with abnormally low production of microglial cells (less than $10^6$ microglial cells per animal) were excluded.
    \item No data had to be excluded as these samples were not used in the study.
    \item The criteria have been thoroughly applied.
\end{enumerate}

\noindent Randomization is not applicable in this study. For details on blinding, outcome measures, statistical methods, experimental animals, and experimental procedures, please refer to the methods section. For information on the results, please refer to the results section.

\bibliography{sample}

\section*{Declarations}
All animal experiments were approved by the institutional animal care and use committee CEEA -- 005 and in agreement with the European guidelines and legislation N°2010/63 UE. The project was approved by the French Ministère de l'Enseignement Supérieur et de la Recherche. The founding males of the C57BL/6J$-$3110043O21Rikem5Lutzy/J (C9orf72 $-/-$) line were provided to us by Jackson Laboratory and those of the B6.129S4 (SJL) $-$ Grntm2.1Far/J (GRNR493X/R493X) line were provided by~\cite{M}. All animal procedures were performed according to the guidelines and regulations of the Institut national de la santé et de la recherche médicale (INSERM) ethical committees (authorization number 2017072018501270). Both males and females were included in the study. Mice were maintained in standard conditions with food and water ad libitum in the ICM animal facilities.

\section*{Acknowledgments}

This study received financial support from the Fondation Alzheimer and was further funded by the 'Investissements d’avenir' program under grant ANR-10-IAIHU-06. The experimental work involving animals was carried out at the PHENO-ICMice core facility. We extend our gratitude to Dr. Julie Smeyers and Elena Gaia Banchi for their invaluable assistance with the mice and primary culture preparations. Additionally, this research was facilitated by the use of equipment and services from the CELIS (Paris Brain Institute, Paris, France), a core facility funded by ANR-10-IAIHU-06 and ANR-11-INBS-0011-NeurATRIS.  This work was granted access to the HPC resources of IDRIS under the allocation 2023-AD011014513 made by GENCI. Lastly, we acknowledge the reviewers whose thoughtful feedback has played a crucial role in enhancing the quality of our manuscript

\section*{Author contributions statement}
OM was responsible for the prototyping, development, and validation of the pipeline modules, including both deep learning and interpretable deep learning components, using the Python programming language. OM also produced qualitative and quantitative results by creating the figures and illustrations featured in this article. LM contributed to the generation of the phagocytosis dataset and provided supervision and validation for the time-lapse methodology. Additionally, LM offered expertise in microglia and Frontotemporal Dementia biology, assisting in the interpretation of the results. RD's role involved supervision and validation of the time-lapse methodology, as well as validation of the image processing pipeline. RD also contributed to the design and validation of the explainable artificial intelligence capabilities, ensuring that the pipeline is both effective and interpretable for users.

\section*{Additional information}
We have made the PhagoStat pipeline publicly available on GitHub. We have also provided a comprehensive dataset, consisting of 235,288 individual files, which have been compressed into a single 94 GB zip file to facilitate downloading. Moreover, we have included detailed statistical reports and the essential source code for generating the genotype comparison figures presented in the article. This approach ensures that readers have complete access to the methodologies utilized and can reproduce the results, thereby augmenting the transparency and reliability of our research findings. 
The repository is accessible using the following link: \url{https://github.com/ounissimehdi/PhagoStat}

\section*{Data availability statement}
All the information and data can be accessed through our repository using the following link: \url{https://github.com/ounissimehdi/PhagoStat} and \url{https://zenodo.org/records/10803492}.

\subsection*{Competing interests}
The authors declare no competing interests.

\newpage
\section{Figures}

\newpage

\newpage

\newpage

\newpage

\newpage

\newpage
\section{Extended Data}\label{extended}

\newpage

\newpage

\newpage

\newpage

\newpage
\begin{sidewaystable}

\begin{tabular*}{\textwidth}
    {@{\extracolsep{\fill}}llccc@{\extracolsep{\fill}}}
    \hline
    & & \multicolumn{2}{@{}c@{}}{Absolute error} & \\
    \cmidrule{3-4}
        & & along x-axis & along y-axis & Execution time (sec)\\
    \hline
    & ECC~\cite{correlation_reg} & 196.02$\pm$119.34 & 206.93$\pm$116.50 & \bfseries 0.92$\pm$0.72 \\
    & SIFT~\cite{SIFT} & \underline{0.0153$\pm$0.0609} & \underline{0.0228$\pm$0.1221} & \underline{$3.09\pm0.12$} \\
    \hline
    \bfseries Ours & \bfseries CECC (n=5) & \bfseries 0.0079 $\pm$ 0.0046 & \bfseries 0.0081$\pm$0.0047 & 8.77$\pm$0.20 \\
    \hline

\end{tabular*}
\caption{Performance evaluation of our CECC registration method compared to the state-of-the-art: We report the results as the mean $\pm$ standard deviation, calculated over 1,000 registration tests. Independent random shifts along the x and y axes were generated within a range of $\pm400$ pixels for $2048\times2048$ pixel images. The best metrics per column are bolted, and the second-best metrics are underlined. Absolute error is calculated based on the difference between the estimated registration coordinates and the ground truth, which are the generated shifts along the x and y axes. Registration time cost is determined by the time taken to register a pair of images (reference and shifted). We demonstrate that ECC is ineffective for the specified registration task, and that the SIFT exhibits a directional bias. In contrast, our proposed CECC (n=5) is unbiased and performs significantly better than both approaches. We conducted the evaluation using the following hardware: a 4-core Xeon Gold 6126 CPU and 1GB RAM. For the SIFT method, we used 2GB RAM, as 1GB was insufficient.}\label{tab:reg}

\begin{center}
\renewcommand{\arraystretch}{1.25}

\begin{tabular*}{23cm}{@{\extracolsep{\fill}}llcccccccc@{\extracolsep{\fill}}}\hline
& & &\multicolumn{4}{@{}c@{}}{Detection metrics} & Semantic metric & \multicolumn{2}{@{}c@{}}{Speed metrics}\\
\cmidrule{3-7}\cmidrule{8-8}\cmidrule{9-10}%
& & mIoU(\%) & F1(\%) & Accuracy(\%) & Precision(\%) & Recall(\%) & Dice(\%) & Train epochs & Inference(sec)\\
\hline
& Cellpose~\cite{cellpose} & $63.85\pm1.39$ & $83.35\pm1.41$& $72.22\pm1.99$& $91.30\pm1.75$ & $77.09\pm1.24$&$84.51\pm1.16$ & 500&$1.032\pm0.0238$ \\
& Stardist~\cite{starconex} &$66.21\pm0.96$& $85.82\pm0.93$ & $76.04\pm1.31$ & \bfseries 94.80$\pm$0.15& $78.95\pm1.47$ & $87.86\pm0.52$ & 400& $0.222\pm0.0173$ \\
\hline
\bfseries Ours & Att-UNet+LSTM & $71.17\pm2.77$ & $86.12\pm2.52$ & $76.73\pm3.54$ & $92.20\pm3.20$ & $81.33\pm2.72$ & $91.87\pm2.6$ & 40 &  $0.369\pm0.0062$\\
& \bfseries Att-UNet (XAI) & \bfseries 73.55$\pm$1.41& \underline{$86.53\pm1.64$} & \underline{$77.00\pm2.47$} & $89.00\pm2.63$ & \bfseries 84.53$\pm$1.55 &  \underline{$93.77\pm0.34$} &  \bfseries 20 &  \underline{$0.157\pm0.0021$}\\

& UNet+LSTM & $72.23\pm1.95$ & \bfseries 86.90$\pm$1.80 & \bfseries 77.72$\pm$2.53 & \underline{$93.47\pm1.61$} & $81.66\pm2.20$ & \bfseries 94.04 $\pm$ 0.28 & 40 &  $0.369\pm0.0062$ \\
& UNet (XAI)& \underline{$72.29\pm2.6$}  & $85.44\pm2.12$ & $75.72\pm2.92$ & $87.95\pm2.87$ & \underline{$83.56\pm2.46$} & $93.18\pm1.18$ & \bfseries 20 & \bfseries 0.136$\pm$0.0069\\
    
& BiONet+LSTM& $63.94\pm6.48$ & $80.25\pm5.41$ & $68.44\pm7.01$ & $90.83\pm2.88$ & $72.73\pm7.44$ & $92.2\pm2.78$ & 40 & $0.306\pm0.0062$\\
& BiONet (XAI) & $65.94\pm5.31$ & $81.43\pm4.13$ & $70.09\pm5.33$ & $88.67\pm3.67$ & $75.99\pm5.95$ & $91.24\pm2.66$ & \bfseries 20 & $0.203\pm0.0042$\\
\hline
\end{tabular*}
\caption{\textbf{Five-fold testing: Quantitative performance evaluation of the cell segmentation module (DL/IDL) compared to state-of-the-art methods.} The reported results are the $(mean \pm standard~deviation)$, computed over 5-fold testing. The best metrics (per column) are highlighted in bold, and the second-best metrics are underlined. Instance-level segmentation (detection) evaluations were used to assess performance with different metrics (per cell mask). The mIoU (mean Intersection over Union) is calculated as the sum of IoU (cell mask-wise) of the predicted cell masks divided by the ground-truth cell count. To report these metrics, we used $IoU\geq50\%$ between ground truth and predicted masks to compute TP, FP, and FN. The $F1$ score is defined as $F1 = \frac{2TP}{2TP+FP+FN}$, while the $Accuracy = \frac{TP}{TP+FP+FN}$, $Precision = \frac{TP}{TP+FP}$, and $Recall = \frac{TP}{TP+FN}$. We utilized a semantic segmentation metric (i.e., Dice coefficient) to quantify the foreground/background pixel-wise separation, defined as $Dice = \frac{2\lvert gt \cap pred \rvert}{\lvert gt \rvert + \lvert pred \rvert}$, where $gt$ is the ground truth mask and $pred$ is the predicted mask ($background = 0, foreground = 1$). The training epochs refer to the number of epochs needed to complete the training phase. Inference time (on the test set) per image was computed using the following hardware: an 8-core i7 9700K CPU, 16GB RAM, NVIDIA MSI 2080}.\label{tab:seg}
\end{center}
\end{sidewaystable}

\newpage
\begin{sidewaystable}

\begin{center}
\renewcommand{\arraystretch}{1.25}

\begin{tabular*}{23cm}{@{\extracolsep{\fill}}llcccccccc@{\extracolsep{\fill}}}\hline
& & &\multicolumn{2}{@{}c@{}}{Cell tracking challenge testing datasets}\\
\cmidrule{2-6}
CTC metrics    & PhC-C2DH-U373        &DIC-C2DH-HeLa        &Fluo-N2DL-HeLa      & Fluo-N2DH-GOWT1   & Fluo-N2DH-SIM+\\
\hline
$OP_{CSB}$(\%) & 95  (CALT-US:96.1)   & 84.3 (CALT-US:92.6) & 86.4 (BFR-GE:95.7) & 88.1 (KTH-SE:95.2) & 84.3 (KIT-GE:90.5)\\
\hline
$SEG$(\%)       & 91.7 (CALT-US:93.1) & 80.1 (CALT-US:88.7) & 78.8 (MU-US:92.3) & 85.2 (CSU-CN:93.8)   & 72.2 (DKFZ-GE:83.2)\\
\hline
$DET$(\%)       & 98.3 (CALT-US:99.0) & 88.5 (CALT-US:97.5) & 94.1 (KIT-GE:99.4) & 91.1 (TUG-AT:98.0)  & 96.5 (FR-GE:98.1)\\

\hline

\end{tabular*}
\caption{\textbf{Quantitative Performance Evaluation of the Cell Segmentation IDL Module on Cell Tracking Challenge Test Datasets\cite{Maska2023}.} This evaluation uses results computed by the challenge organizers\cite{Maska2023} upon submission of our test set results (where no ground truth was provided). The AttUnet(XAI) performance was assessed using the challenge metrics against the test ground truth masks. The $OP_{CSB}$ metric is defined as $0.5 \times (DET + SEG)$. Here, $DET = 1 - \frac{\min(AOGM-D, AOGM-D_{0})}{AOGM-D_{0}}$, which represents the normalized Acyclic Oriented Graph Matching (AOGM-D) measure for detection\cite{Matula2015}. The SEG measure employs the Jaccard similarity index, expressed as $J(S, R) = \frac{|R \cap S|}{|R \cup S|}$, where $R$ is the set of pixels in a reference object and $S$ is the set of pixels in its matching segmented object. A reference object $R$ and a segmented object $S$ are considered matching if $|R \cap S| > 0.5 \times |R|$. The top-performing apporach per dataset and per metric are mentioned as follow (approach : score) from CTC website\cite{Maska2023}.}\label{tab:TESTCTCseg}
\end{center}
\end{sidewaystable}

\end{document}